\documentclass[trackchanges]{aastex7}

\shorttitle{DR21SF}
\shortauthors{Yang et al.}

\begin{document}

\title{Unveiling Fiber Networks and Core Formation in the DR21 South Filament}

\author[orcid=0000-0002-9839-185X]{Kai Yang}
\affiliation{School of Astronomy and Space Science, Nanjing University, 163 Xianlin Avenue, Nanjing 210023, People's Republic of China}
\affiliation{Key Laboratory of Modern Astronomy and Astrophysics (Nanjing University), Ministry of Education, Nanjing 210023, People's Republic of China}
\affiliation{Department of Astronomy, School of Physics and Astronomy, and Shanghai Key Laboratory for Particle Physics and Cosmology, Shanghai Jiao Tong University, Shanghai 200240, People's Republic of China}
\affiliation{State Key Laboratory of Dark Matter Physics, Shanghai Jiao Tong University, Shanghai 200240, People's Republic of China}
\email{kyang@nju.edu.cn}

\author[orcid=0000-0002-5093-5088]{Keping Qiu}
\affiliation{School of Astronomy and Space Science, Nanjing University, 163 Xianlin Avenue, Nanjing 210023, People's Republic of China}
\affiliation{Key Laboratory of Modern Astronomy and Astrophysics (Nanjing University), Ministry of Education, Nanjing 210023, People's Republic of China}
\email[show]{kpqiu@nju.edu.cn}

\author[orcid=0000-0003-1337-9059]{Xing Pan}
\affiliation{School of Astronomy and Space Science, Nanjing University, 163 Xianlin Avenue, Nanjing 210023, People's Republic of China}
\affiliation{Key Laboratory of Modern Astronomy and Astrophysics (Nanjing University), Ministry of Education, Nanjing 210023, People's Republic of China}
\affiliation{Center for Astrophysics $\vert$ Harvard \& Smithsonian, 60 Garden Street, Cambridge, MA 02138, USA}
\email{xing.pan@cfa.harvard.edu}

\begin{abstract} \label{abs}

We present high-resolution ($\sim$1000 AU) 3~mm observations with the NOrthern Extended Millimeter Array toward the DR21 South Filament, aiming to reveal its internal fragmentation and search for deeply embedded star-forming activities. 
Both the continuum and molecular line emissions align well with the filament axis traced by the low-resolution ($\sim$18$^{\prime\prime}$) column density map. The 3~mm continuum, CS (2$-$1), and HCO$^+$ (1$-$0) emissions reveal continuous and diffuse structures with measured FWHM widths of 0.054, 0.029, and 0.030 pc, respectively. In contrast, the H$^{13}$CO$^+$ (1$-$0) emission appears more clumpy and localized. 
The non-thermal motion in the filament is predominantly subsonic to transonic.
We detect 13 dense cores in NH$_2$D (1$_{11}-1_{01}$), three of which coincide with continuum peaks; virial analysis suggests most are gravitationally bound. 
Using a friend-of-friend algorithm, we identify 32, 34, and 22 velocity-coherent fibers from the CS, HCO$^+$, and H$^{13}$CO$^+$ data, respectively. Compared to fibers traced by CS and HCO$^+$, H$^{13}$CO$^+$ fibers are more frequently associated with NH$_2$D cold cores and exhibit higher average mass-per-unit-length values.
Differences among CS, HCO$^+$, and H$^{13}$CO$^+$ emissions likely arise from variations in effective critical densities.
These results are consistent with a hierarchical structure, in which the 3.6-pc DR21SF contains velocity-coherent fibers and gravitationally bound dense cores.

\end{abstract}


\section{Introduction} \label{intro}

Galactic-wide surveys conducted by the \textit{Spitzer} and \textit{Herschel} Space Telescopes have uncovered the presence of massive filamentary clouds distributed across the Milky Way \citep{2009PASP..121..213C,2010A&A...518L.100M}. These elongated structures, spanning several to over ten parsecs, are often not isolated but instead embedded within more complex structures, such as interconnected webs or clustered filament systems (e.g., \citealt{2011A&A...529L...6A}; \citealt{2013ApJ...764L..26B}). In many cases, they form radial networks that converge toward dense central regions, commonly referred to as hub-filament system (HFS, e.g., \citealt{2009ApJ...700.1609M}; \citealt{2012A&A...540L..11S}; \citealt{2018A&A...613A..11W}).

Accumulating observations indicate that these filaments are not monolithic structures but are composed of intricate internal structures. A pioneering study by \citet{2013A&A...554A..55H} revealed the existence of ``velocity-coherent structures'', commonly known as fibers. Since then, similar substructures have been reported in both low- and high-mass star-forming filaments (e.g., \citealt{2018A&A...610A..77H}; \citealt{2022ApJ...927..106C}; \citealt{2024A&A...687A.140H}). However, our understanding of fibers remains limited, and there are some key questions: (1) What are the physical properties of these filamentary substructures? (2) What role do they play in shaping the initial conditions for star formations? Addressing these questions requires observations of filaments that are relatively quiescent and largely unaffected by star formation feedback, as that their internal structures remain undisturbed.
In this context, the DR21 South Filament (DR21SF), a massive and quiescent filament, stands out to be a valuable laboratory for exploring the intrinsic nature of filamentary substructures.

\begin{figure}[h!]
   \includegraphics[width=175mm]{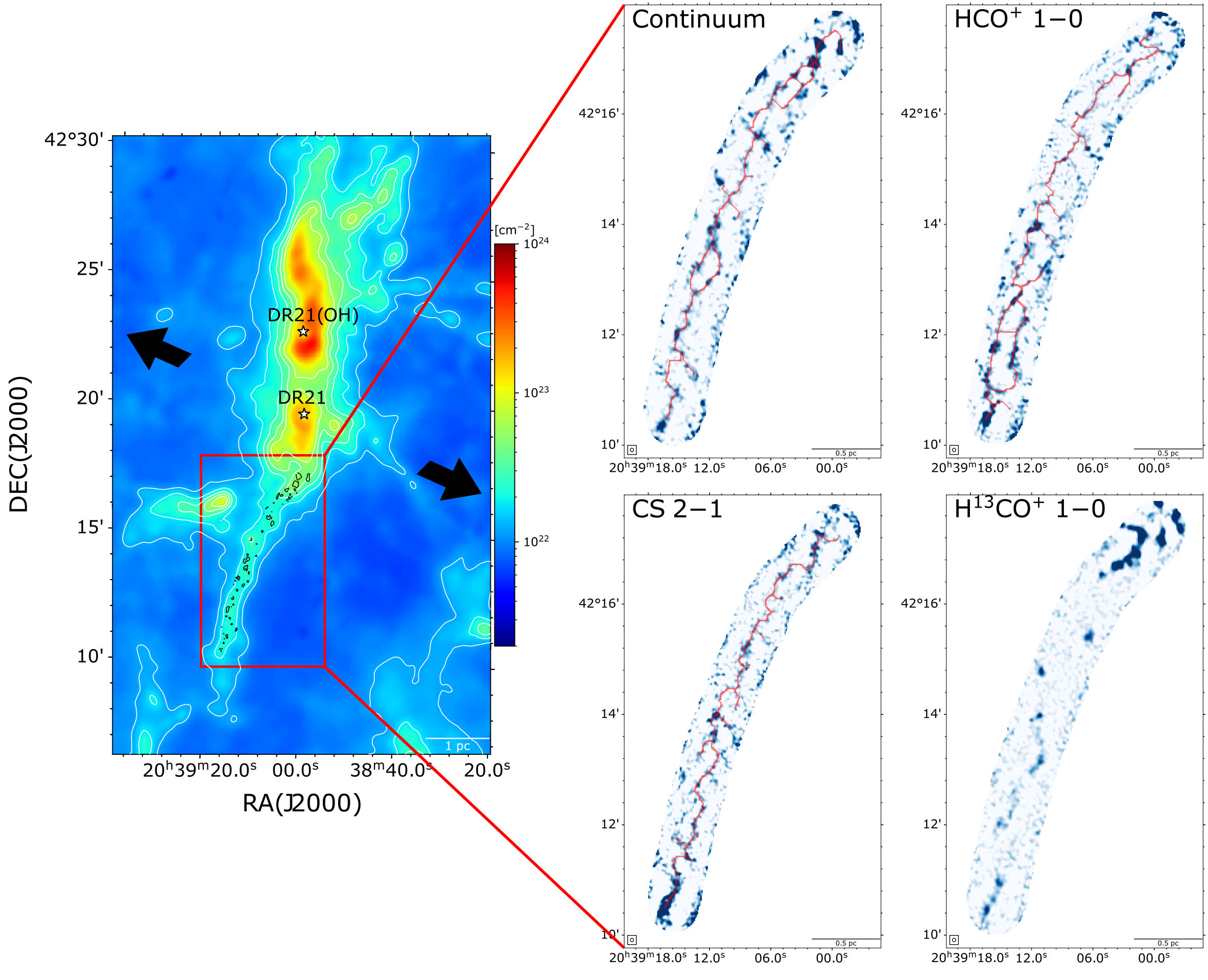}
     \caption{Column density, continuum, and molecular line maps of DR21SF. Left: Column density map of the DR21 filament obtained from \citet{2019ApJS..241....1C}. White contours indicate levels of $1.5^{[0,1,2,3,\dots]}\times1.2\times10^{22}$ cm$^{-2}$. Black contours mark the 3$\sigma$ level of the 3~mm continuum emission. The two well-known massive dense cores, DR21 and DR21(OH), are indicated by stars. Black arrows show the orientation of the DR21 outflow. Right (four panels): Continuum, HCO$^{+}$, CS, and H$^{13}$CO$^{+}$ moment-0 maps. Molecular line emissions are integrated over the velocity range from $-6.4$ to $0.8$ km s$^{-1}$. Identified filamentary skeletons are overlaid in red. The synthesized beam is shown in the lower-left corner of each panel.}
  \label{fig:2d_fila}
\end{figure}

Located within the Cygnus-X complex, the DR21 filament is one of the most massive and actively star-forming filaments in this region \citep{2007A&A...476.1243M,2010A&A...520A..49S,2011ApJ...740L...5C,2022ApJ...927..106C}, situated approximately 1.4 kpc from the Sun \citep{2012A&A...539A..79R}. As shown in the left panel of Figure~\ref{fig:2d_fila}, the main DR21 filament appears as a bright ridge to the north, and harbors two massive dense cores, DR21 and DR21(OH). DR21 coincides with a group of luminous H\,{\sc ii}\, regions, and drives the most luminous ($\sim$ 1800 $L_{\odot}$) and massive ($\geq$ 3000 $M_{\odot}$) outflow in the Galaxy (known as the DR21 outflow, \citealt{1986MNRAS.220..203G}; \citealt{2007MNRAS.374...29D}). \citet{2012A&A...543L...3H} further identified two filaments extending from this outflow center: the S filament and the SW filament. The more prominent of the two, DR21SF, extends $\sim$8$^{\prime}$ to the south, spanning 3.6 pc with a total mass of 1048 $M_{\odot}$ \citep{2021ApJ...908...70H}. In contrast to other dense gas filaments in the region which exhibit clear signs of ongoing massive star formation (e.g., W75N, \citealt{2010A&A...520A..49S}; \citealt{2021ApJ...918L...4C}), DR21SF appears remarkably quiescent and presents a smooth column density profile with few star formation activity \citep{2014AJ....148...11K}, making it an ideal environment for studying internal filamentary substructures without significant contamination from star formation feedback processes.

In this study,  we investigate the internal structure of DR21SF by mapping the filament in 3mm continuum and multiple molecular line transitions using the NOrthern Extended Millimeter Array (NOEMA). The observational setup and data reduction procedures are described in Section\ref{obs}. In Section~\ref{sec:res}, we identify the velocity-coherent fibers and characterize their physical properties. Section~\ref{dis} discusses the differences among molecular tracers and examines the role of fibers as intermediate structures linking large-scale filaments and star-forming cores. A summary of our main findings is provided in Section~\ref{Sum}.


\section{Observations} \label{obs}

We carried out a 16-pointing mosaic toward the DR21 South Filament in 2018 with NOEMA. The overall observing time is 10 hours. The observations were carried out in D configuration with baselines between 15m and 175m. The DR21SF was observed in the 3~mm continuum emission and in four spectral units covering the NH$_{2}$D (1$_{11}-1_{01}$), H$^{13}$CO$^{+}$ (1$-$0), HCO$^{+}$ (1$-$0), and CS (2$-$1) lines at 85.93, 86.75, 89.19, and 97.98 GHz, respectively. We observed the quasar 2013-370 and the evolved star MWC349 for phase and flux calibration, respectively. The phase centre for the observations of DR21SF is $\alpha$(J2000) = 20$^{\rm h}$39$^{\rm m}$09$\fs834$, $\delta$(J2000) = 42$\degr$14$'$23$\farcs$98.

Data reduction, calibration, and imaging were performed with the \texttt{CLIC} and \texttt{MAPPING} programs of the \texttt{GILDAS}\footnote{\url{http:/www.iram.fr/IRAMFR/GILDAS}} software package developed by the IRAM and Observatoire de Grenoble. The continuum was extracted from the line-free channels with a synthesized beam of 4$\farcs6$ $\times$ 3$\farcs8$, position angle (P.A.) of 2.8$\degr$, and an rms (root-mean-square) noise level of 40 $\mu$Jy beam$^{-1}$. For the molecular spectral lines, we achieve a synthesized beam of 4$\farcs4$ $\times$ 3$\farcs8$ (P.A. = 10.5$\degr$) and an rms noise level of 20 mJy beam$^{-1}$ per 65.1 kHz (corresponding to 0.22 km s$^{-1}$) channel.


\section{Results and analysis} \label{sec:res}

DR21SF is mostly dark at infrared wavelengths to 70 $\mu m$ indicating its early evolutionary stage. Figure \ref{fig:2d_fila} presents an overview of DR21SF in column density. The column density map is adopted from \cite{2021ApJ...918L...4C}, derived through spectral energy distribution (SED) fitting of Herschel PACS (160 $\mu m$) and SPIRE (250, 320, and 500 $\mu m$) images, with an angular resolution of 18$^{\prime\prime}$. DR21SF is apparent in the column density map, extending from DR21 to the southeast with an inclination angle of $\sim$18$^{\circ}$ \citep{2021ApJ...908...70H}.

\subsection{Dust and molecular line emissions}

The 3~mm continuum emission is overlaid as contours on the column density map in the left panel of Figure \ref{fig:2d_fila}. Although the 3$\sigma$ continuum contours are fragmented, they align well with the filament skeleton. 
In addition, the right panels of Figure \ref{fig:2d_fila} present the 3~mm continuum map and the integrated intensity maps of HCO$^{+}$, CS, and H$^{13}$CO$^{+}$. 
The continuum, as well as the HCO$^{+}$ and CS emissions, exhibits relatively continuous and diffuse structures, while the H$^{13}$CO$^{+}$ emission appears more clumpy. Despite this difference, all tracers show emission distributed along the filamentary structure.

To trace the crests of the most prominent filamentary structures within the DR21SF region, we employed the \texttt{DisPerSE} algorithm \citep{2011MNRAS.414..350S}, which utilizes discrete Morse theory to analyze the topological features of 2D datasets. \texttt{DisPerSE} has been widely applied in filament identification  (e.g., \citealt{2011A&A...529L...6A}; \citealt{2012A&A...540L..11S}; \citealt{2014prpl.conf...27A}; \citealt{2014MNRAS.444.2507P}; \citealt{2018A&A...610A..62C}; \citealt{2019A&A...623A.142S}; \citealt{2022A&A...664A..88E}). To obtain the most visually optimal and successive filamentary structures, the data were first smoothed to a 18$^{\prime\prime}$ resolution, equivalent to the angular resolution of the column density map, before apply \texttt{DisPerSE}. We adopted persistence thresholds of 1.5$\times$10$^{-4}$ Jy beam$^{-1}$, 0.2, and 0.2 km s$^{-1}$ Jy beam$^{-1}$ for the 3~mm continuum, HCO$^{+}$, and CS emissions, respectively. Adjacent skeletons with an orientation difference of less than 70$^{\circ}$ were merged into one. The filaments identified by \texttt{DisPerSE} are shown as red lines in Figure \ref{fig:2d_fila}. The filament, as revealed by different tracers, is apparent with a roughly northwest-southeast orientation, coincident with the filamentary structure reported in \citet{2012A&A...543L...3H}. In addition, several short, branch-like filaments merge into the main filaments, and ring-like structures are found in both the continuum and HCO$^{+}$ maps. The skeletons obtained from different tracers show a broadly similar morphology but differ in detail, which may be related to optical depth effects (with the continuum being optically thin while the two lines are moderately optically thick) or to variations in the critical densities traced by different transitions.

To measure the Full Width at Half Maximum (FWHM) of the filaments, we extracted perpendicular intensity profile along each filament and fitted the median radial profiles using a Gaussian function (details provided in Appendix \ref{app1}). The resulting FWHMs for the continuum, HCO$^{+}$, and CS filaments are 0.054, 0.029, and 0.030 pc, respectively. 
These widths are comparable to those of the Serpens South molecular cloud ($\sim$0.035 pc; \citealt{2014ApJ...790L..19F}), Orion ($\sim$0.02 $-$ 0.05 pc for OMC-1/2 and ISF; \citealt{2018A&A...610A..77H}), L1287 ($\sim$0.03 pc; \citealt{2020A&A...644A.128S}), and Barnard 5 ($\sim$0.03 pc; \citealt{2021ApJ...909...60S}),
but narrower than the width of DR21SF (0.13 pc) as derived from the column density map by \citet{2021ApJ...908...70H}, as well as the ``universal" 0.1 pc filament width proposed for nearby star-forming cloud by \citet{2011A&A...529L...6A}. However, given the angular resolution of our observations ($\sim$4.1$^{\prime\prime}$, corresponding to 0.028 pc at a distance of 1.4 kpc), the measured widths are likely not fully resolved. This is because part of the more extended emission is filtered out in the NOEMA data, so that we mainly detect the narrower central ``spines'' of the filament, resulting in FWHM values that are close to the beam size.


\begin{figure}[h!]
   \includegraphics[width=175mm]{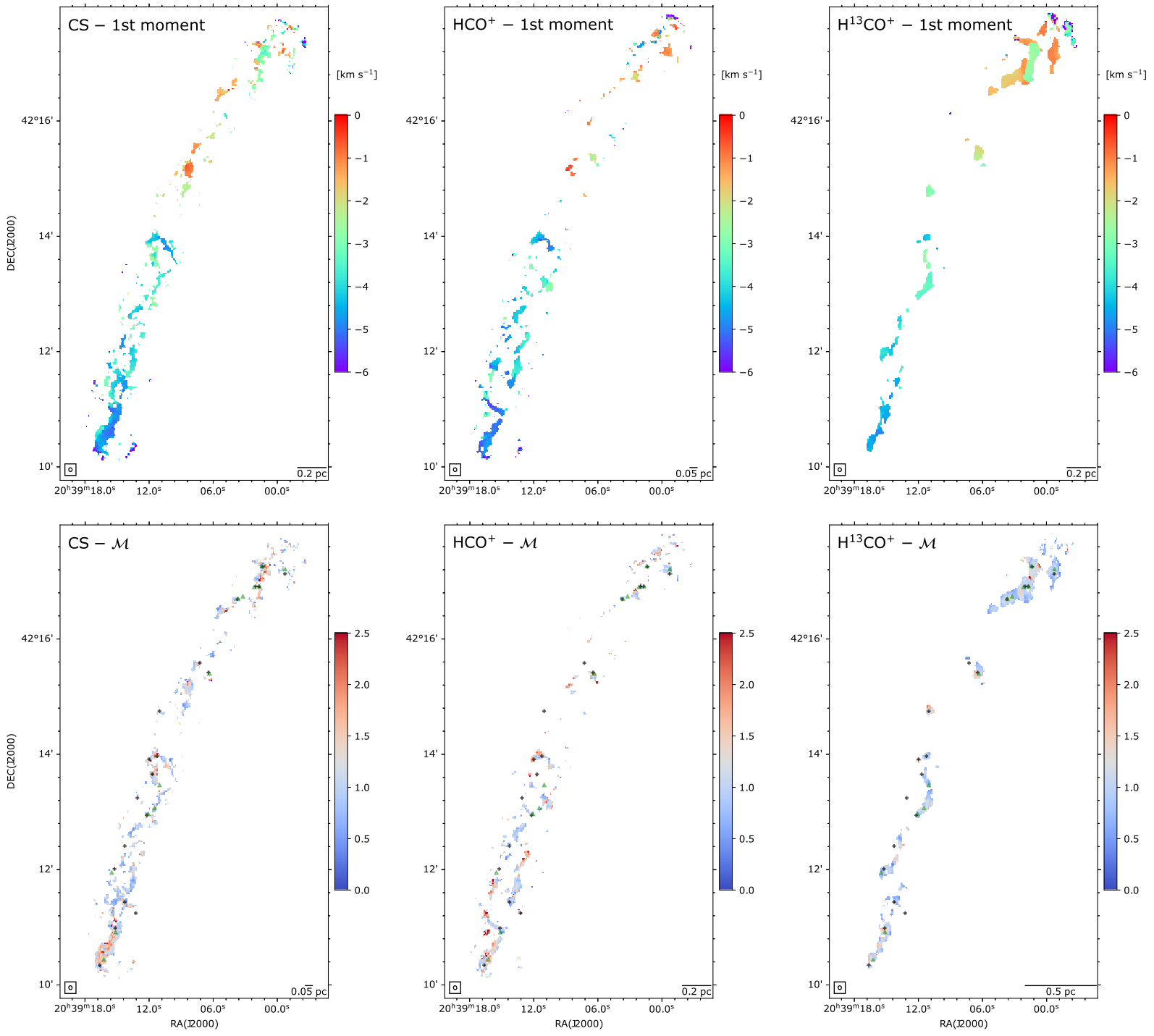}
     \caption{Distributions of the central velocity (top three panels) and Mach number (bottom three panels) derived from the HCO$^{+}$, CS, and H$^{13}$CO$^{+}$ lines in DR21SF. Black crosses mark the continuum peak positions with intensities higher than 5$\sigma$ and the green triangles indicate the positions of the NH$_{2}$D cores. The synthesized beam is shown in the bottom-left corner and the scale bar is presented in the bottom-right corner.}
  \label{fig:v0_mach}
\end{figure}

\subsection{Velocity structures and Mach numbers} \label{sec:fit}

We perform pixel-by-pixel Gaussian fitting of molecular line image cubes (CS, HCO$^{+}$, and H$^{13}$CO$^{+}$), using the \texttt{PySpecKit} package \citep{2011ascl.soft09001G}. Firstly, we focus on regions with the peak intensity in the spectral domain exceeding five times the RMS noise level, and take spectral peaks that are separated by more than three channels and brighter than five times the RMS noise level as independent velocity components. We then fit Gaussian profiles to these components based on the following criteria: (1) the central velocity, $v_{\rm lsr}$, needs to be in the range of [$-$6, 0] km s$^{-1}$, corresponding to the velocity range in the DR21SF from \citep{2021ApJ...908...70H}; (2) the observed velocity dispersion, $\sigma_{\rm obs}$, must be larger than three times its uncertainty to ensure its reliability; (3) the integrated intensity must exceed three times its uncertainty. 

In the Gaussian fitting, single velocity component was resolved in most (\textgreater~70\%) of the areas where significant molecular line emissions were detected (Table \ref{tab_mach}). We present the fitted central velocity of these line emissions in Figure \ref{fig:v0_mach}. They reveal a consistent velocity gradient along the filament's north-south axis for all three molecular tracers, which is also reported in \citet{2021ApJ...908...70H}. Among the tracers, CS and HCO$^{+}$ show broader spatial coverage and finer velocity structures compared to H$^{13}$CO$^{+}$, which is more confined to dense substructures. This suggests that H$^{13}$CO$^{+}$ preferentially traces denser and more compact regions, whereas CS and HCO$^{+}$ probe more extended gas.

Based on the fitted observational velocity dispersion, $\sigma_{\rm obs}$, which consists of thermal and nonthermal components, the Mach number is estimated as $\mathcal{M}$ = $\sigma_{\rm nt}$/$c_{\rm s}$, where $\sigma_{\rm nt}$ = $\sqrt{\sigma_{\rm obs}^{2}-(\Delta_{\rm ch}/2 \sqrt{2 \rm ln2})^{2}-\sigma_{\rm th}^{2}}$ is the nonthermal velocity dispersion, and $c_{\rm s}$ is the sound speed. The channel width $\Delta_{\rm ch}$ = 0.22 km s$^{-1}$, the molecular thermal velocity dispersion $\sqrt{k_{\rm B}T/\mu m_{p}}$, where $k_{\rm B}$ is the Boltzmann constant, $T$ is assumed to be 15 K based on \citet{2019ApJS..241....1C}. $\mu$ is the molecular weight, with values of 44, 29, and 30 for CS, HCO$^{+}$, and H$^{13}$CO$^{+}$ respectively, and $m_{p}$ is the proton mass. 

Figure \ref{fig:v0_mach} shows the spatial distributions of Mach numbers derived from the Gaussian fitting to the CS, HCO$^{+}$ and H$^{13}$CO$^{+}$ line emissions. In these maps, the majority of the observed regions exhibit subsonic or transonic nonthermal motions (0 $<$ $\mathcal{M}$ $\leq$ 2). The Mach numbers calculated based on HCO$^{+}$ range from 0.02 to 5.40, with a median value of 1.08, closely matching those from CS (0.03 to 4.51, median 1.02). The H$^{13}$CO$^{+}$ emission yields a slightly lower median Mach number of 0.97 (see Table \ref{tab_mach} for detailed statistics). Over 90\% of the filament is subsonic ($\mathcal{M}$ $\leq$ 1) or transonic (1 $<$ $\mathcal{M}$ $\leq$ 2) in nonthermal motions, confirming the quiescent nature of DR21SF. Furthermore, no significant enhancement of Mach numbers is found toward the continuum peaks or NH$_{2}$D cores, suggesting that these cold dense cores are also quiescent and is not associated with active star formation. 
These findings emphasize that DR21SF is a dense, relatively undisturbed filament with minimal influence from star formation feedback. 
Although star formation is already taking place within it, the overall environment remains quiescent, making DR21SF an ideal target for investigating the initial conditions of filament fragmentation and the onset of star formation.

\begin{figure}[h!]
   \includegraphics[width=88mm]{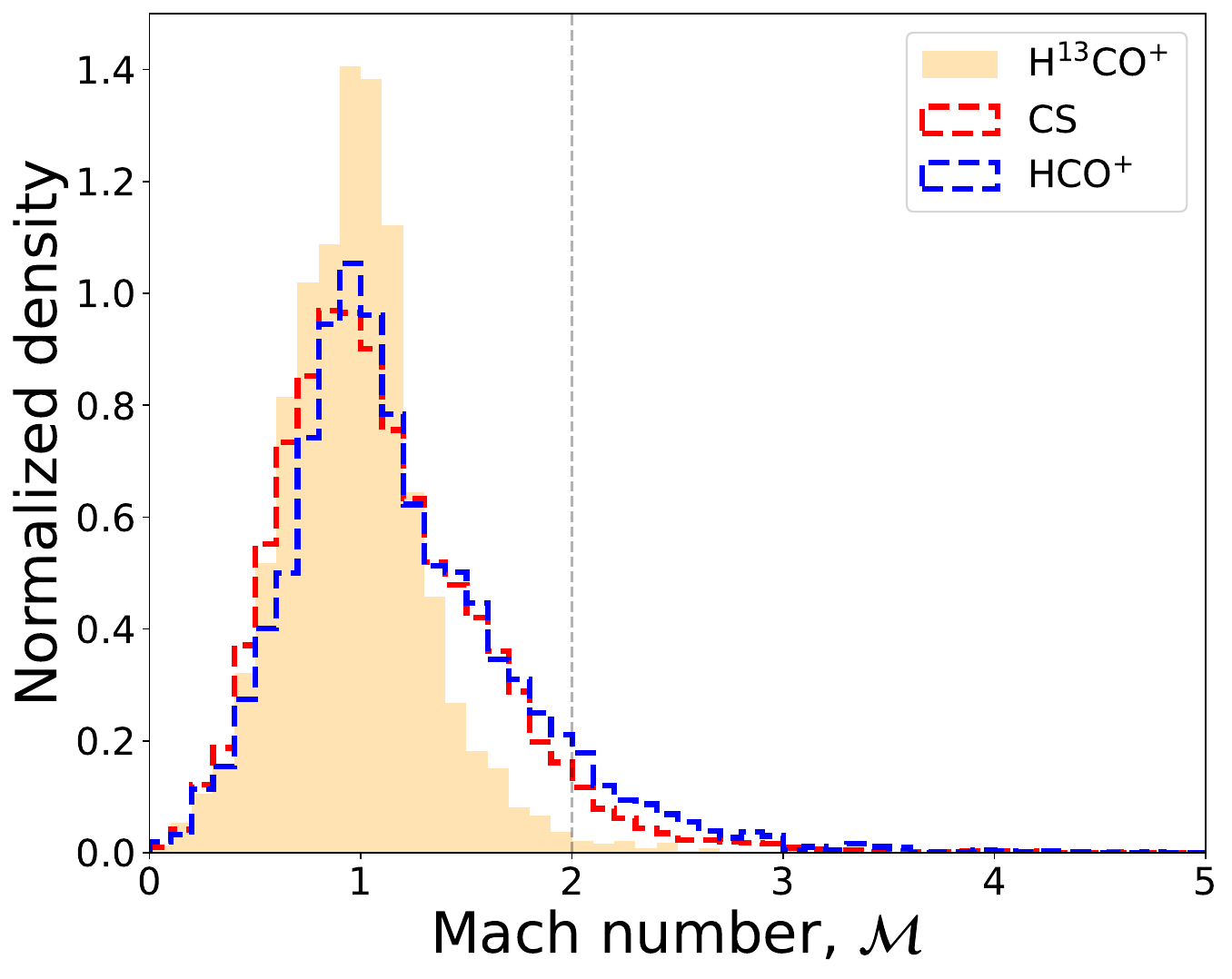}
   \includegraphics[width=88mm]{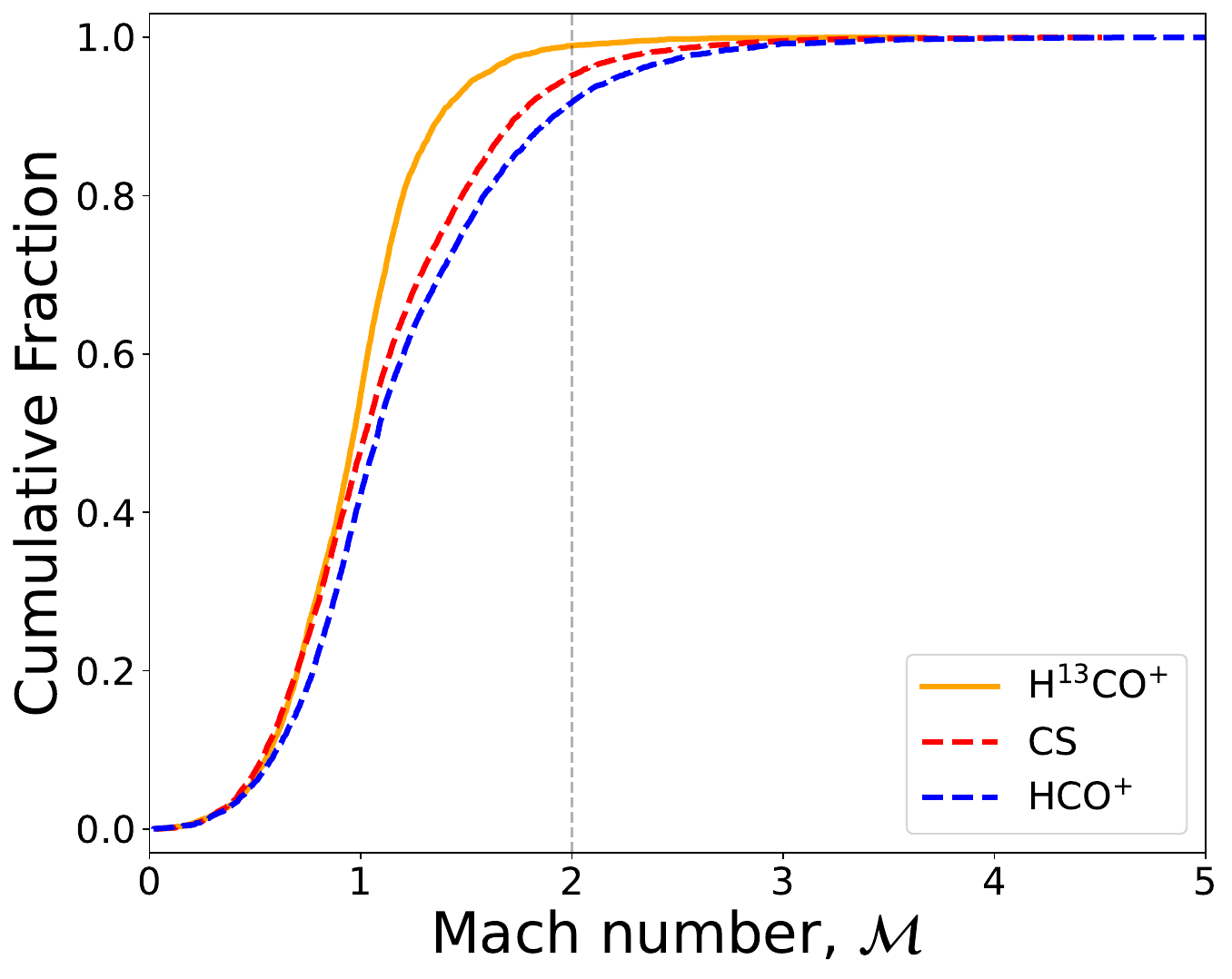}
     \caption{The Mach number distribution for the HCO$^{+}$, CS, and H$^{13}$CO$^{+}$ emissions. Left: normalized probability distributions of the Mach numbers. Right: Cumulative distributions of the Mach numbers. The gray vertical dashed line represents the position of the Mach number equal to 2.}
  \label{fig:mach}
\end{figure}

Figure \ref{fig:mach} presents the normalized histogram (the left panel) and cumulative distributions (the right panel) of the Mach numbers. While the CS and HCO$^{+}$ lines display similar Mach number distributions, H$^{13}$CO$^{+}$ shows a narrower distribution around $\mathcal{M} = 1$. The implications of these differences will be discussed in Section \ref{dis:eff_cri}. \citet{2021ApJ...908...70H} reported significantly higher Mach numbers based on NH$_3$ observations with a beam size of 48$^{\prime\prime}$. 
Using the measured velocity gradient of 0.8 s$^{-1}$ pc$^{-1}$ inferred from our high-resolution data, we estimate that unresolved gradients would contribute a linewidth of only $\sim$0.26 km s$^{-1}$ within a 48$^{\prime\prime}$ beam. In comparison, NH$_3$ shows a median intrinsic linewidth of 1.3 km s$^{-1}$. Therefore, unresolved velocity gradients can only account for a minor fraction of the broader NH$_3$ lines, indicating that intrinsic nonthermal motions are the dominant contributor to the higher Mach numbers.

\begin{table*}
\movetableright=-0.3in
\caption{Mach Number}  \label{tab_mach}
\begin{tabular}{c c c ccc ccc}
\hline\hline
Molecules & Pixel & Single & $\mathcal{M}$ & $\mathcal{M}_{\rm mean}$ & $\mathcal{M}_{\rm median}$ & $\mathcal{M}$ $\leq$ 1 & 1 $<$ $\mathcal{M}$ $\leq$ 2 & $\mathcal{M}$ $>$ 2 \\
 & Numbers & Component & & & & & & \\
(1) & (2) & (3) & (4) & (5) & (6) & (7) & (8) & (9) \\
\hline
CS        & 7740 & 71.0\% & 0.03$-$4.51 & 1.10 & 1.02 & 48.0\% & 47.2\% & 4.8\% \\
HCO$^{+}$ & 5639 & 75.6\% & 0.02$-$5.40 & 1.19 & 1.08 & 42.3\% & 49.5\% & 8.2\% \\
H$^{13}$CO$^{+}$ & 5372 & 78.7\% & 0.05$-$4.66 & 0.98 & 0.97 & 54.9\% & 44.0\% & 1.1\% \\
\hline\hline
\end{tabular}
\end{table*}

\subsection{Velocity-coherent fibers resolved from the NOEMA observation}
\subsubsection{Fiber identification}

Based on the Gaussian fitting results described in Section \ref{sec:fit}, we employed a Python-based friend-of-friend (FoF) algorithm\footnote{\url{https://github.com/ShanghuoLi/pyfof}} \citep{1982ApJ...257..423H} to identify velocity-coherent fibers. The procedure follows the approach outlined in \citet{2013A&A...554A..55H}, \citet{2018A&A...610A..77H}, and \citet{2022ApJ...926..165L}, and consists of the following steps:
\begin{enumerate}
    \item We first select seed points with peak intensities greater than 7$\sigma$, which represent approximately 72\% of the data points. The FoF algorithm is then applied to group these points based on the following criteria: (a) the spatial separation between neighboring points must be $\Delta r \leq 0.025$ pc, equivalent to one synthesized beam size; (b) the velocity gradient threshold is defined as $\nabla v_{\rm LSR} = \Delta v / 2\theta_{\rm beam}$, following \citet{2018A&A...610A..77H}, where $\Delta v$ is the spectral line’s FWHM and $\theta_{\rm beam}$ is the beam size; (c) each group must consist of at least 60 points, corresponding to three times the beam area size.
    \item The remaining points with intensities between 5$\sigma$ and 7$\sigma$ are then considered. The FoF algorithm is rerun to associate these lower-intensity points with the previously identified groups using the same spatial and kinematic criteria. Through these two steps, over 80\% of the fitted data points are successfully grouped into velocity-coherent structures.
    \item To trace the central axes of the identified fibers, we used the \texttt{FilFinder}\footnote{\url{http://github.com/e-koch/FilFinder}} package, which applies the Medial Axis Transform to extract the skeletons of filamentary structures. 
\end{enumerate}

\begin{figure}[h!]
   \includegraphics[width=175mm]{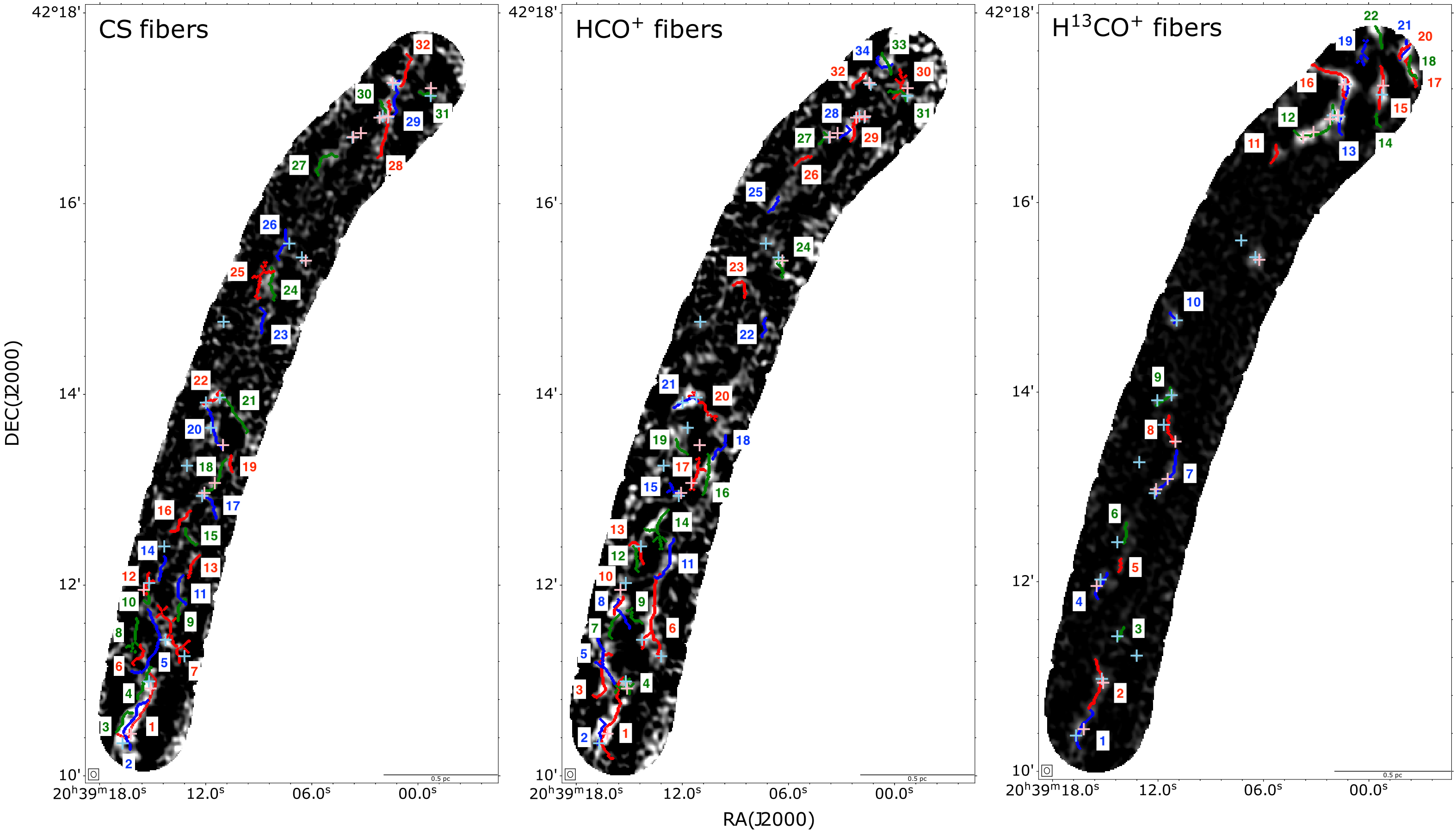}
     \caption{Identified fibers along the DR21SF filaments based on CS, HCO$^{+}$, and H$^{13}$CO$^{+}$ line emissions. The red, green, and blue segments represent the extracted fiber spines, overlaid on the corresponding molecular integrated intensity maps. Continuum peak positions with intensity higher than 5$\sigma$ are marked with sky-blue crosses, and NH$_2$D core positions are denoted by cyan crosses.}
  \label{fig:fiber}
\end{figure}

A total of 32, 34, and 22 velocity-coherent fibers are identified for the CS, HCO$^{+}$, and H$^{13}$CO$^{+}$ lines, respectively. Their spines are shown as red, green, and blue segments in Figure~\ref{fig:fiber}. Overall, the CS and HCO$^{+}$ maps exhibit more extensive and densely populated fibers compared to the H$^{13}$CO$^{+}$ map. There is a notable spatial coincidence between many fiber spines and continuum peaks or NH$_{2}$D cold cores, especially in the CS and HCO$^{+}$ panels. These results imply that the DR21SF is composed of multiple entangled fiber-like substructures, consistent with the ``filament bundle" scenario proposed by \citet{2013A&A...554A..55H}. Additional information and individual properties of these fibers can be found in Appendix \ref{app2}.

Figure~\ref{fig:fiber_vlsr} presents the $V_{\rm LSR}$ position-velocity diagram of fibers associated with NH$_{2}$D cores. Some fibers are excluded because they lack velocity association with the cores despite their spatial overlap. CS and HCO$^{+}$ fibers tend to have similar $V_{\rm LSR}$ distributions, indicating that they are probing similar kinematic structures. In contrast, H$^{13}$CO$^{+}$ fibers display a clear offset in the velocity space when compared to CS and HCO$^{+}$, and the NH$_{2}$D cores are frequently found to be associated with H$^{13}$CO$^{+}$ fibers. In addition, we grouped fibers from different tracers into the same matched set when they exhibit overlapping spatial distributions and similar centroid velocities, with detailed information provided in Appendix \ref{app2}. In total, 19 matched sets are identified. CS and HCO$^{+}$ share a substantial number of these sets (15 in total), whereas only seven H$^{13}$CO$^{+}$ fibers are matched with fibers traced in CS or HCO$^{+}$. Notably, three matched sets contain fibers detected in all three tracers.

\begin{figure}[h!]
   \includegraphics[width=180mm]{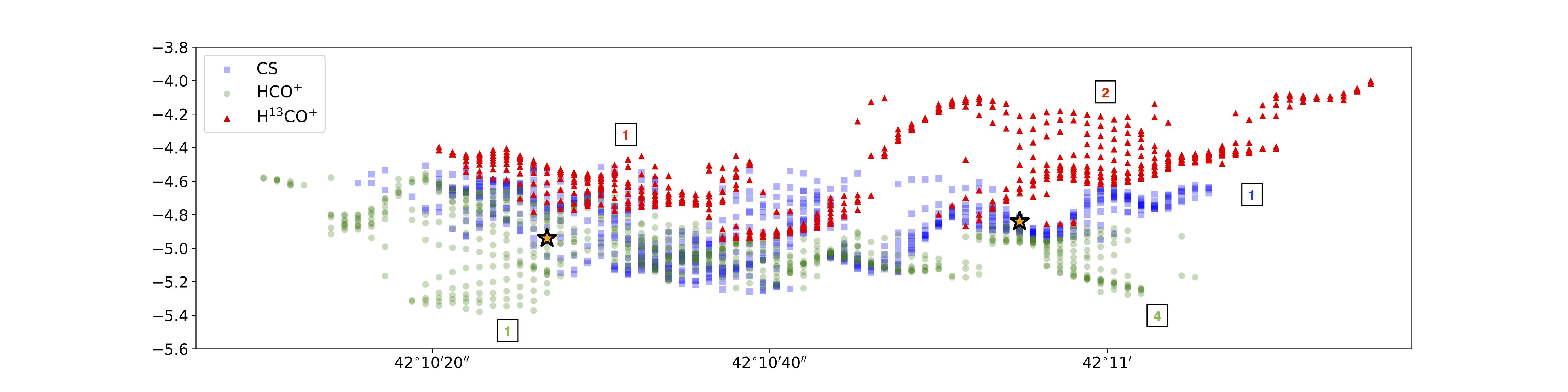}\\
   \includegraphics[width=180mm]{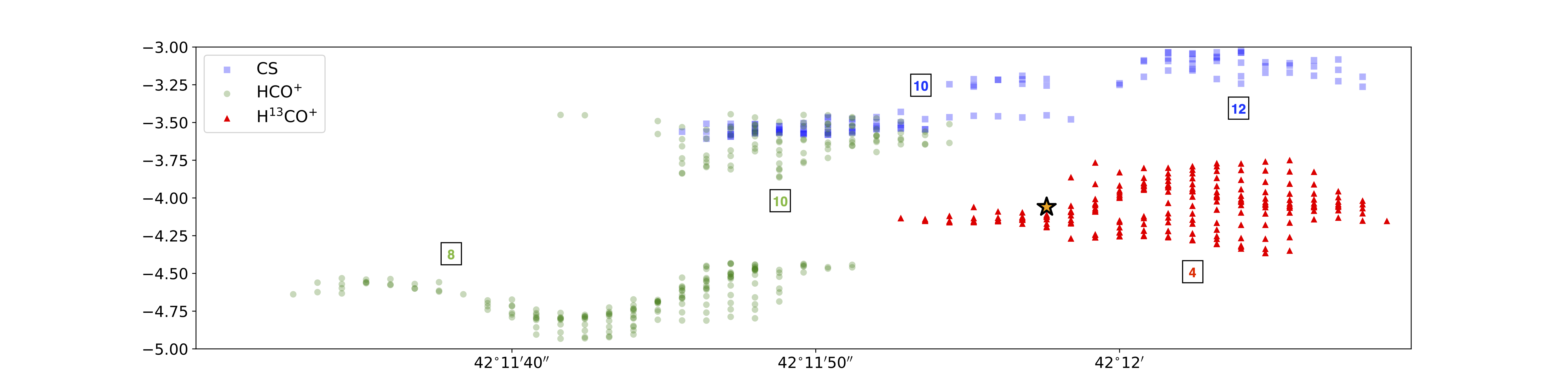}\\
   \includegraphics[width=180mm]{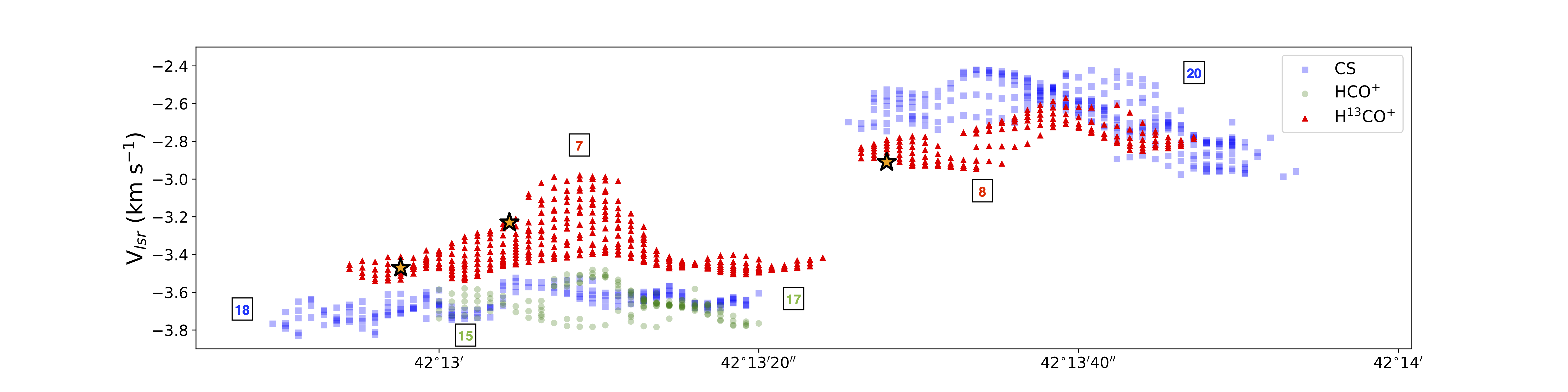}\\
   \includegraphics[width=180mm]{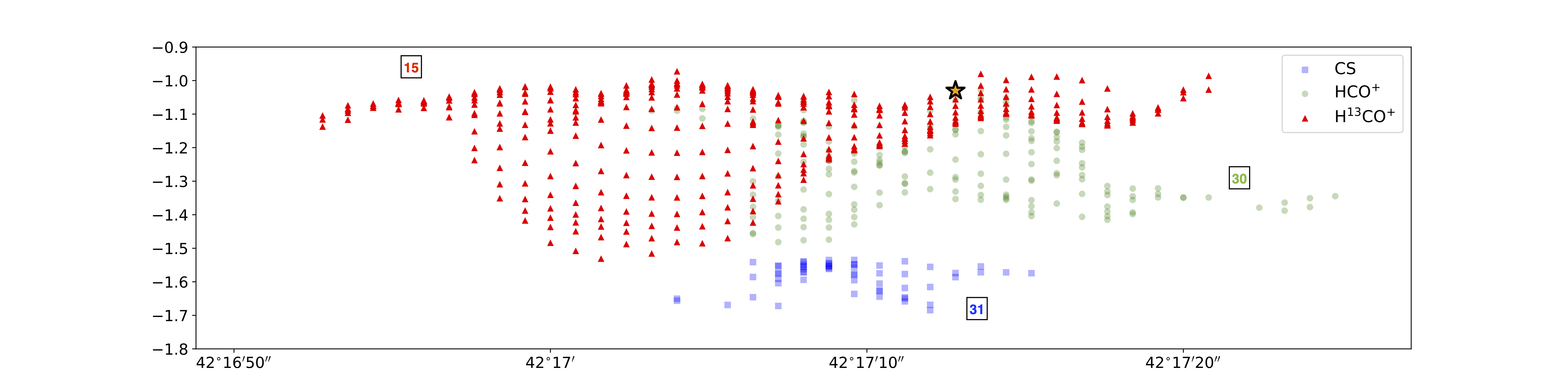}\\
   \includegraphics[width=180mm]{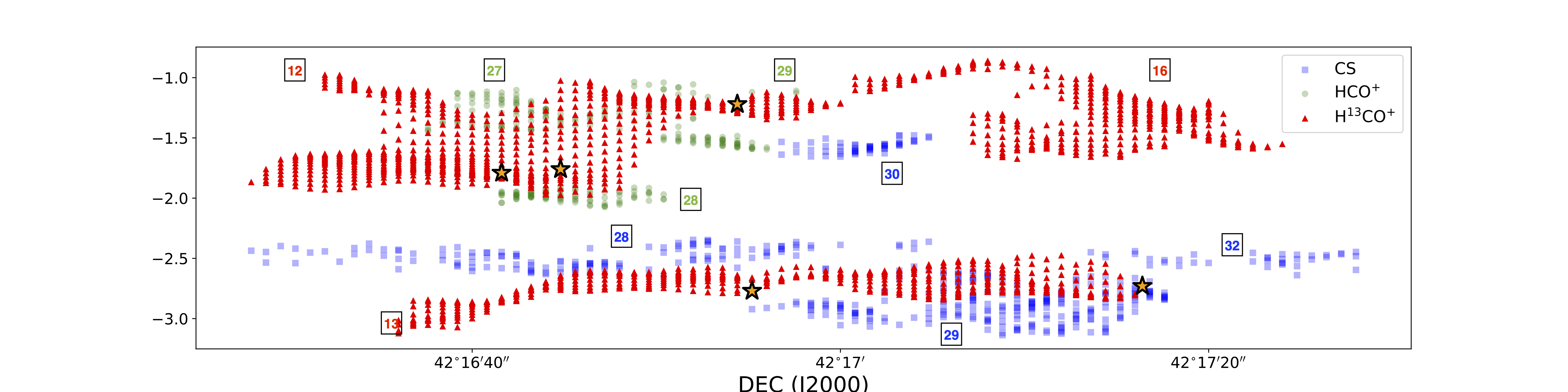}
     \caption{Velocity profiles of fibers associated with NH$_{2}$D cores along the declination axis. Blue, green, and red squares are CS, HCO$^{+}$, and H$^{13}$CO$^{+}$ data respectively. The stars represent the NH$_{2}$D cores. From top to bottom, the five panels are arranged in spatial order from south to north.}
  \label{fig:fiber_vlsr}
\end{figure}

\subsubsection{Fiber profiles}

\begin{figure}[h!]
   \includegraphics[width=175mm]{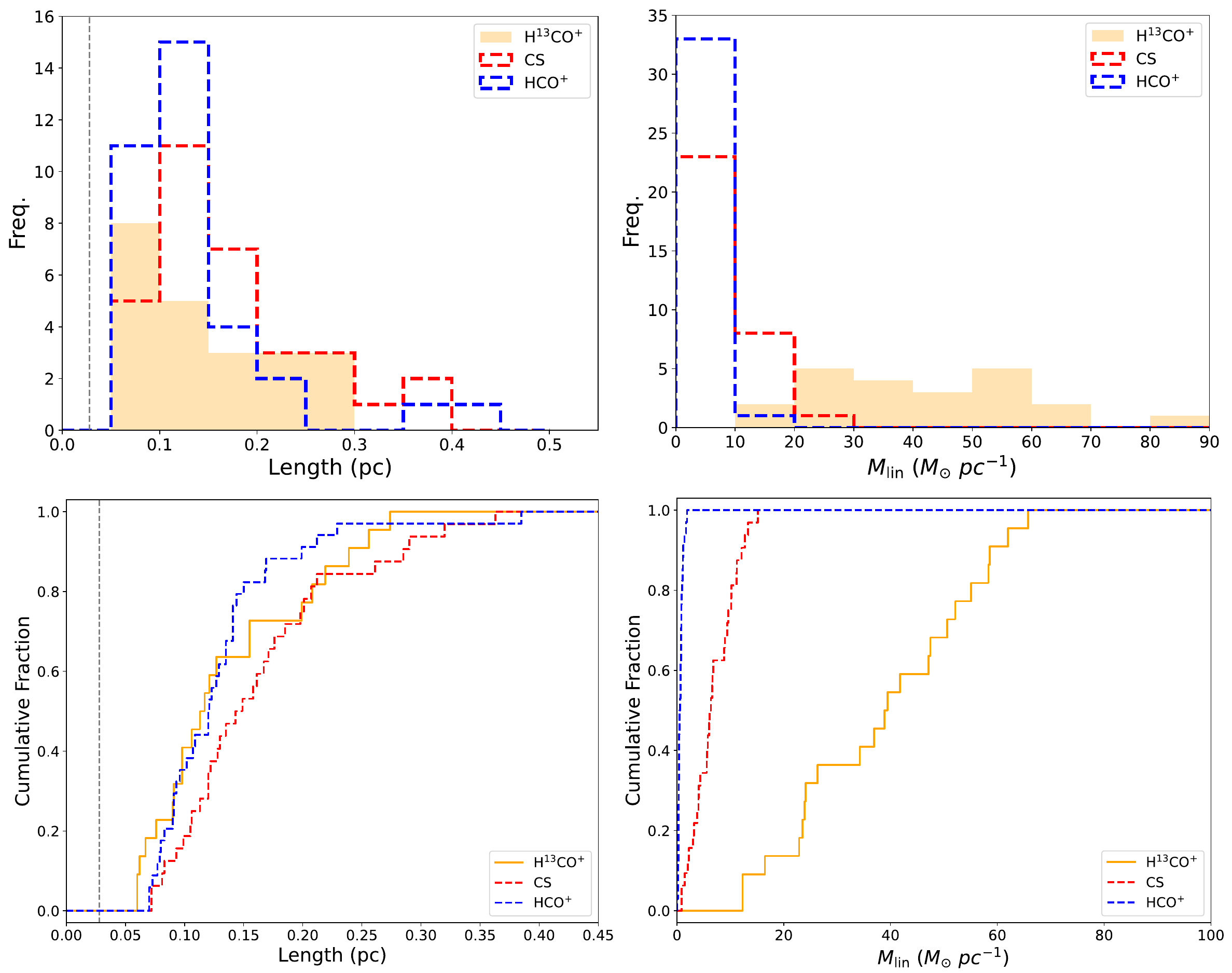}
     \caption{Statistical properties of the fibers. Left: fiber length, the dashed lines represent the beam size. Right: mass-per-unit-length. The properties derived from CS, HCO$^{+}$, and H$^{13}$CO$^{+}$ are highlighted by red, blue, and orange, respectively.}
  \label{fig:fiber_prop}
\end{figure}

The spine lengths of these fibers ($L_{\rm fiber}$) range from 0.07$–$0.39 pc for CS, 0.07$–$0.42 pc for HCO$^{+}$, and 0.06$–$0.29 pc for H$^{13}$CO$^{+}$, and their statistical distributions are shown in Figure~\ref{fig:fiber_prop}. The CS and HCO$^{+}$ fiber lengths have broader ranges than the H$^{13}$CO$^{+}$ fibers, extending to $\sim$0.4 pc, and they both exhibit peaks between 0.1$-$0.15 pc. The H$^{13}$CO$^{+}$ fibers are relatively shorter, with a peak lower than 0.1 pc. These fiber lengths are comparable to those of fibers detected in the Orion Integral-shaped Filament \citep{2018A&A...610A..77H}.

The mass-per-unit-length is a key parameter for characterizing the stability of fibers. With the measured fiber length, it is desirable to calculate the fiber mass, which is usually measured by the continuum emission. However, the continuum emission is not significantly (most of them $<$ 5$\sigma$) and continuously detected (Figure \ref{fig:2d_fila}), we instead utilize the molecular emissions to estimate the fiber masses. The fiber mass is estimated using the following formula:
\begin{equation}\label{fila_mass}
M_{\rm fiber} = \mu_{\rm H_{2}}~m_{\rm H}~\Sigma \frac{N(\rm molecule)}{X(\rm molecule)}~\Omega ,\\
\end{equation}
where $\mu_{\rm H_{2}} = 2.8$ is the mean particle weight per H$_{2}$ molecule \citep{2008A&A...487..993K}, $m_{\rm H}$ is the mass of a hydrogen atom, $\Omega$ is the solid angle subtended by the molecular line emission, $N({\rm molecule})$ is the column density, and $X({\rm molecule})$ is the fractional abundance of the molecule. We adopt $X({\rm molecule})$ values of $10^{-9}$ for CS, $3 \times 10^{-9}$ for HCO$^{+}$, and $6 \times 10^{-11}$ for H$^{13}$CO$^{+}$, based on \citet{2014A&A...563A..97G} and \citet{2019MNRAS.488.2357P}.
We note that the HCO$^{+}$ and CS emissions may be optically thick and interferometric observations are less sensitive to extended and relatively diffuse emissions. Consequently, the masses derived from HCO$^{+}$ and CS are likely to represent lower limits to the fiber mass.

Assuming local thermodynamic equilibrium (LTE) conditions and optically thin line emission, the column densities can be derived following \citet{2015PASP..127..299S}:
\begin{equation}\label{Eq_N}
N_{\rm tot} = \left(\frac{3h}{8 \pi^{3} S \mu^{2} R_{i}}\right) \left(\frac{Q(T_{\rm ex})}{g_{u}}\right) \frac{{\rm exp}\left(\frac{E_{u}}{kT_{ex}}\right)}{{\rm exp}\left(\frac{h\nu}{kT_{ex}}\right)-1} \times \frac{1}{(J_{\nu}(T_{ex})-J_{\nu}(T_{bg}))} \int \frac{T_{R}dv}{f} , \\
\end{equation}
where $h$ is the Planck constant, $S$ is the line strength for linear molecules, $\mu$ is the permanent dipole moment, $R_i = 1$ for $\Delta J = 1$ transitions, $Q(T_{\rm ex})$ is the partition function, $g_u$ is the statistical weight of the upper level, $E_u$ is the energy of the upper level, $k$ is the Boltzmann constant, $T_{\rm ex}$ is the excitation temperature, $\nu$ is the rest frequency of the transition, $\int T_R dv$ is the integrated intensity, $f$ is the beam filling factor (assumed to be 1), $J_\nu(T)$ is the Planck function, and $T_{\rm bg} = 2.73$~K is the cosmic microwave background temperature. The excitation temperature is taken from the dust temperature map in \citet{2019ApJS..241....1C}. Spectroscopic parameters are obtained from the CDMS \citep{2001A&A...370L..49M,2005JMoSt.742..215M,2016JMoSp.327...95E} and JPL \citep{1998JQSRT..60..883P} molecular catalogs.

The resulting mass-per-unit-length, defined as $M_{\rm line} = M_{\rm fiber} / L_{\rm fiber}$, span the ranges: 0.9$–$29.5~$M_{\odot}$pc$^{-1}$ for CS fibers, 0.2$–$1.9~$M_{\odot}$pc$^{-1}$ for HCO$^{+}$ fibers, and 12.3$–$85.3~$M_{\odot}$pc$^{-1}$ for H$^{13}$CO$^{+}$ fibers. Their distributions are shown in Figure~\ref{fig:fiber_prop}. The CS and HCO$^{+}$ fibers both show mass-per-unit-length peaks below 10 $M_{\odot}$pc$^{-1}$, whereas the H$^{13}$CO$^{+}$ fibers exhibit significantly higher and more widely distributed $M_{\rm line}$ values.

\subsection{Cold and Quiescent Cores} \label{sec:nh2d}

\subsubsection{Identification of NH$_{2}$D cores}

The NH$_{2}$D (1$_{11}-1_{01}$) transition, with a critical density of $\sim 10^{5}$ cm$^{-3}$ \citep{2022ApJ...929...13C}, is an effective tracer of the dense interior regions of pre-stellar cores \citep{2017A&A...600A..61H}. To identify cold cores, we first preselect regions where the NH$_{2}$D velocity-integrated intensity exceeds the 3$\sigma_{\rm area}$ noise level. We then apply the \texttt{astrodendro}\footnote{\url{http://dendrograms.org}} algorithm \citep{2008ApJ...679.1338R} to extract compact NH$_{2}$D structures. The identification is performed using the following criteria: a minimum intensity threshold of 3$\sigma_{\rm area}$ and a minimum of 20 connected pixels per structure, approximately corresponding to the area size of the synthesized beam.

\begin{figure}[h!]
   \includegraphics[width=175mm]{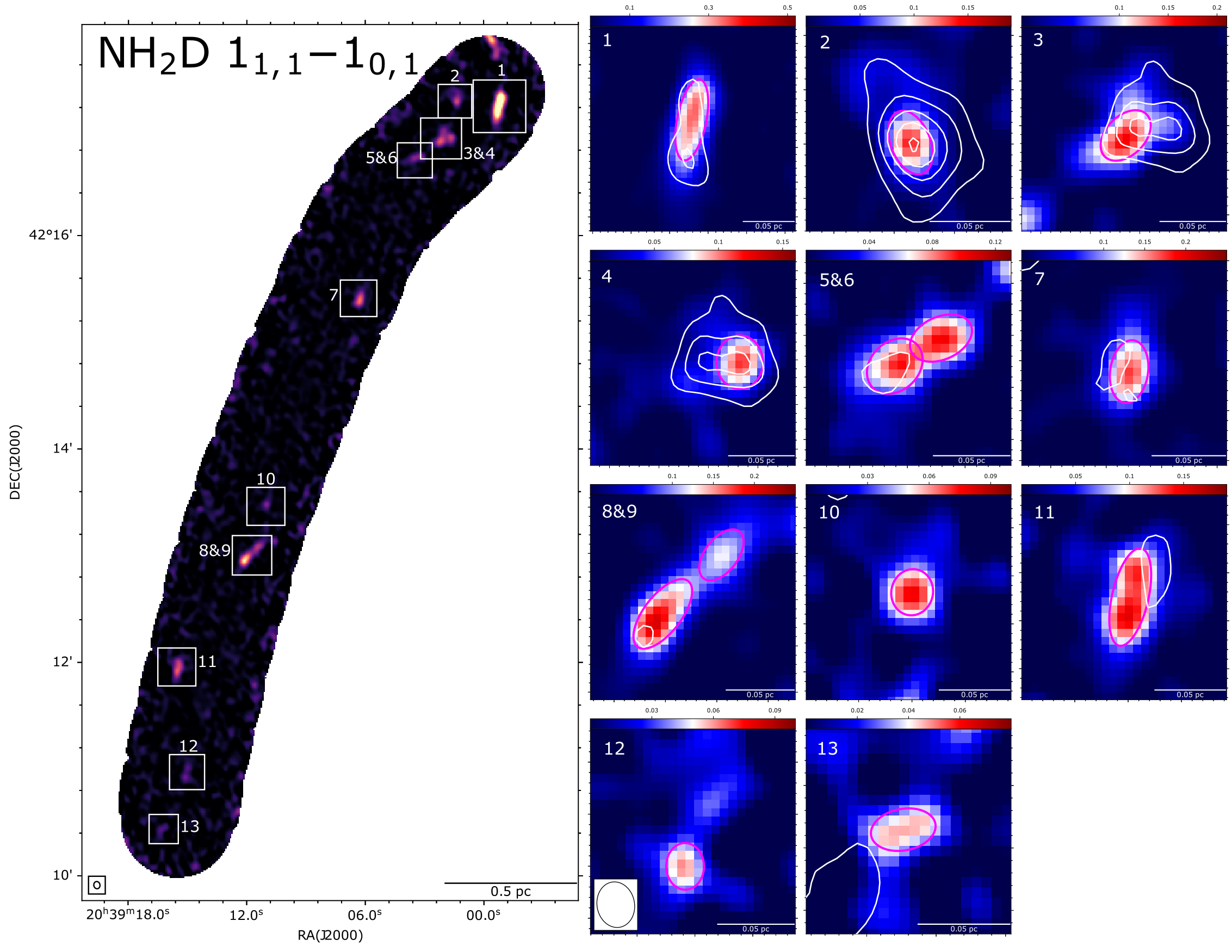}
     \caption{NH$_2$D cores in the DR21SF region. Left: NH$_2$D moment-0 map integrated over the velocity range from $-10.2$ to $5.0$ km s$^{-1}$. Right (11 panels): Identified NH$_2$D cold cores. Each color map displays the NH$_2$D moment-0 emission integrated over a $\pm$1.2 km s$^{-1}$ range centered on the systemic velocity of each core. Pink ellipses indicate the FWHM of the identified NH$_2$D cores. White contours represent the continuum emission at levels of (3, 6, 9, 12) $\times$ $\sigma$.}
  \label{fig:nh2d}
\end{figure}

As a result, thirteen compact structures are identified in the NH$_{2}$D emission. Their integrated intensity maps are shown in Figure \ref{fig:nh2d}, and their physical properties (e.g., position and size) are summarized in Table \ref{tab_nh2d}. Among these, four NH$_{2}$D cores are spatially coincident with 3 mm continuum peaks, four show clear offsets between their NH$_{2}$D and continuum peaks, and the remaining five exhibit no detectable continuum counterparts above the 3$\sigma$ level. Based on systematic review of publicly available data on Orion A, \citet{2016A&A...590A...2S} proposed that the evolved objects may be moving away from their natal dense filament due to a slingshot mechanism, implying that the more evolved cores tend to be found further away from the filament. Based on this, we assume a temperature of T = 10 K for the four NH$_{2}$D cores that are well aligned with the continuum peaks, and T = 15 K for the four with offset positions, when estimating their gas masses and virial parameters.

\subsubsection{Gas masses and dynamical states}

Assuming optically thin dust emission, we estimate the gas mass ($M_{\rm gas}$) by adopting the equation:
\begin{equation}\label{equ:gas_mass}
M_{\rm gas} = \eta \frac{S_{\nu} d^{2}}{B_{\nu}(T) \kappa_{\nu}} ,\\
\end{equation}
where $\eta$ = 100 is the assumed gas-to-dust mass ratio, $d$ is the source distance of 1.4 kpc, $S_{\nu}$ is the continuum flux at frequency $\nu$, $B_{\nu}(T)$ is the Planck function at temperature $T$, and $\kappa_{\nu}$ is the dust opacity coefficient at frequency $\nu$. We adopt the dust opacity $\kappa_{\nu}$ = 10($\nu$/1.2 THz)$^{\beta}$ cm$^{2}$g$^{-1}$ and $\beta$ = 1.5 following \citet{1983QJRAS..24..267H}.

We can estimate the virial parameter, $\alpha_{\rm vir}$, using the following equation \citep{1992ApJ...395..140B}:
\begin{equation}\label{virial_para}
\alpha_{\rm vir} = \frac{5 \sigma_{v}^{2}R}{G M_{\rm gas}} ,\\
\end{equation}
where $\sigma_{v}$ is the velocity dispersion, $R$ is the radius of the structure, and $G$ is the gravitational constant. $\sigma_{v}$ is estimated as:
\begin{equation}\label{velo_disp}
\sigma_{v} = \sqrt{\sigma_{\rm nt,NH_{2}D}^{2}+c_{s}^{2}} = \sqrt{\sigma_{\rm obs,int}^{2}-\frac{k_{B}T}{\mu m_{p}}+\frac{k_{B}T}{\mu_{p}m_{p}}}  ,\\
\end{equation}
where $c_{s}$ is the isothermal sound speed, $\mu$ = 18 is the molecular weight of NH$_{2}$D, $\mu_{p}$ = 2.37 is the mean molecular weight \citep{2008A&A...487..993K}, and $m_{p}$ is the proton mass. $\sigma_{\rm obs}$ is derived by fitting the core-averaged spectrum with \texttt{PySpecKit}. 

The derived virial parameters, ranging from 0.69 to 4.57, are listed in Table \ref{tab_nh2d}. Most of the NH$_{2}$D cores with reliable mass estimates exhibit virial parameters below 2, suggesting that they are gravitationally bound and potentially undergoing collapse. 
For four cores (\#1, \#7, \#8 and \#11), the 3~mm continuum peaks show clear spatial offsets from the NH$_{2}$D peaks. Within the NH$_{2}$D core boundaries, the continuum emission is not fully or reliably detected (particularly for cores \#8 and \#11), which introduces large uncertainties into the gas mass estimates. Therefore, for these cores the derived masses should be regarded as lower limits, and the corresponding virial parameters are likely overestimated. As a result, their virial parameters exceeding 2 do not imply that the cores are in an expanding state \citep{2013ApJ...779..185K}.
In these estimates, we assume a constant radial density profile and neglect the influence of external pressure.

\begin{table*}
\centering
\caption{Physical Parameters of the NH$_{2}$D Cores.}  \label{tab_nh2d}
\footnotesize
\begin{tabular}{c cc ccc cc ccc cc}
\hline\hline
CoreID & R.A. & Decl. & Maj $\times$ Min & P.A. & R & $\sigma_{\rm obs}$ & V$_{\rm LSR}$ & S$_{\nu}$ & T$_{\rm dust}$ & M$_{\rm gas}$ & $\alpha_{\rm vir}$ & T$_{\rm vir}$ \\
 & (hh:mm:ss) & (hh:mm:ss) &  &  &  &  &  &  &  &  &  & \\
 & (J2000) & (J2000) & ($^{\prime\prime}$ $\times$ $^{\prime\prime}$) & (deg) & (pc) & (km s$^{-1}$) & (km s$^{-1}$) & (mJy) & (K) & (M$_{\odot}$) &  & (K) \\
\hline
\noalign{\smallskip}
 1  & 20:38:59.247 & 42:17:12.687 & 11.9 $\times$ 4.2 & 78.7  & 0.024 & 0.210 & $-$1.03 & 2.17  & 19.0 & 2.26  & 1.00 & 15 \\
 2  & 20:39:01.415 & 42:17:15.998 &  6.1 $\times$ 3.6 & 111.7 & 0.016 & 0.255 & $-$2.73 & 2.14  & 18.7 & 2.26  & 0.69 & 10 \\
 3  & 20:39:02.156 & 42:16:54.160 &  6.7 $\times$ 4.4 & 45.5  & 0.018 & 0.260 & $-$1.22 & 1.68  & 18.6 & 1.79  & 1.24 & 15 \\
 4  & 20:39:01.662 & 42:16:54.953 &  6.0 $\times$ 5.2 & 90.2  & 0.019 & 0.230 & $-$2.77 & 0.786 & 18.6 & 0.840 & 1.95 & 10 \\
 5  & 20:39:03.648 & 42:16:41.928 &  5.9 $\times$ 4.7 & 44.4  & 0.018 & 0.188 & $-$1.79 & 0.587 & 18.7 & 0.620 & 1.90 & 10 \\
 6  & 20:39:03.215 & 42:16:44.404 &  6.3 $\times$ 4.1 & 24.2  & 0.017 & 0.167 & $-$1.76 & $-$   & $-$  & $-$   & $-$  & $-$ \\ 
 7  & 20:39:06.342 & 42:15:24.181 &  6.3 $\times$ 3.8 & 81.8  & 0.016 & 0.186 & $-$2.10 & 0.596 & 20.5 & 0.568 & 1.89 & 15 \\
 8  & 20:39:12.076 & 42:12:58.027 &  9.0 $\times$ 4.0 & 52.4  & 0.020 & 0.206 & $-$3.47 & 0.389 & 19.5 & 0.392 & 3.87 & 10 \\
 9  & 20:39:11.490 & 42:13:04.312 &  6.4 $\times$ 3.6 & 52.8  & 0.016 & 0.230 & $-$3.23 & $-$   & $-$  & $-$   & $-$  & $-$ \\
 10 & 20:39:11.025 & 42:13:28.114 &  4.8 $\times$ 2.3 & 78.2  & 0.016 & 0.317 & $-$2.91 & $-$   & $-$  & $-$   & $-$  & $-$ \\
 11 & 20:39:15.503 & 42:11:57.030 & 10.4 $\times$ 3.8 & 77.1  & 0.021 & 0.209 & $-$4.06 & 0.431 & 19.5 & 0.435 & 4.57 & 15 \\
 12 & 20:39:15.127 & 42:10:54.980 &  4.5 $\times$ 3.6 & 90.8  & 0.014 & 0.178 & $-$4.84 & $-$   & $-$  & $-$   & $-$  & $-$ \\
 13 & 20:39:16.241 & 42:10:26.514 &  5.2 $\times$ 3.3 & 10.5  & 0.014 & 0.245 & $-$4.94 & $-$   & $-$  & $-$   & $-$  & $-$ \\
\hline\hline
\end{tabular}
\end{table*}

\section{Discussion} \label{dis}

\subsection{Differences in molecular line emissions and their implications for tracing dense gas}  \label{dis:eff_cri}

In Section \ref{sec:res}, we presented NOEMA observations of three molecular transitions toward DR21SF in the Cygnus-X complex: CS (2$-$1), HCO$^+$ (1$-$0), and H$^{13}$CO$^+$ (1$-$0). A comparative analysis of these lines, along with the properties of the identified velocity-coherent fibers, may reveal differences in their spatial distributions, kinematic features, and physical conditions traced.

Although CS and HCO$^+$ exhibit similar properties, the H$^{13}$CO$^+$ emission shows difference in several aspects: (1) Morphology: CS and HCO$^+$ show extended and relatively weak emission (typically $\sim$5$\sigma_{\rm area}$) along the main axis of DR21SF, where as H$^{13}$CO$^+$ displays more compact structures confined to smaller regions. 
Notably, the number of fibers identified in CS and HCO$^+$ is comparable (32 and 34), while only 22 fibers are found in H$^{13}$CO$^+$. 
(2) Turbulence: The Mach number derived from CS and HCO$^+$ spans a wider range, indicating overwhelmingly sub- to transonic, with a small fraction of supersonic. H$^{13}$CO$^+$, however, predominantly traces gas with Mach numbers near 1, suggesting more quiescent conditions. 
(3) Velocity structure: 
Most CS and HCO$^+$ fibers fall into 15 matched sets based on their overlapping spatial distributions and similar centroid velocities. In comparison, only seven H$^{13}$CO$^+$ fibers match with CS or HCO$^+$ fibers. Several H$^{13}$CO$^+$ fibers (e.g., \#4, \#6, \#7, and \#11) overlap spatially with CS or HCO$^+$ fibers but exhibit significantly different velocity ranges, suggesting that H$^{13}$CO$^+$ traces distinct kinematic components.
Notably, NH$_2$D cores are more frequently associated with H$^{13}$CO$^+$ structures. 
(4) Physical properties: The median fiber lengths derived from CS, HCO$^+$, and H$^{13}$CO$^+$ are similar (0.15, 0.12, and 0.12 pc, respectively). In contrast, H$^{13}$CO$^+$ fibers tend to be more massive, with median values of 5.0 $M_{\odot}$ and 41 $M_{\odot}~\rm pc^{-1}$, than CS and HCO$^+$ fibers (1.0 and 0.07 $M_{\odot}$; 6.5 and 0.6 $M_{\odot}~\rm pc^{-1}$), supporting its role in highlighting denser gas components. The fiber widths are similar among the three tracers (0.029, 0.027, and 0.030 pc for CS, HCO$^+$, and H$^{13}$CO$^+$, respectively), all close to the beam size of $\sim$4.1$^{\prime\prime}$ ($\sim$0.028 pc at 1.4 kpc).

These observed differences are intriguing given the fact that all three transitions have comparable critical densities. However, a more diagnostic parameter in this context is the effective excitation density ($n_{\rm eff}$), which accounts for optical depth effects and radiative trapping. Defined as the density required to produce an integrated line intensity of 1 K km s$^{-1}$ under typical conditions of column density and kinetic temperature \citep{2015PASP..127..299S}, $n_{\rm eff}$ can vary significantly even among transitions with similar $n_{\rm crit}$.

Specifically, both CS (2$-$1) and HCO$^+$ (1$-$0) are sensitive to radiative trapping due to their relatively higher optical depths, resulting in effective excitation densities that are 1–2 orders of magnitude lower than their critical densities. In contrast, the optically thin H$^{13}$CO$^+$ (1$-$0) line has an $n_{\rm eff}$ that is much closer to its $n_{\rm crit}$, and consequently, significantly higher than those of CS and HCO$^+$ (see Table 1 in \citet{2015PASP..127..299S}). This makes H$^{13}$CO$^+$ more sensitive in tracing high-density gas of great relevance to star formation, whereas CS and HCO$^+$ are more sensitive to extended, lower-density envelopes and filamentary structures.

This interpretation is consistent with findings from other regions. For example, \citet{2024A&A...687A.140H} reported that HC$_3$N (10$–$9) emission, with an $n_{\rm eff}$ of 4.3 $\times$ 10$^5$ cm$^{-3}$, appears more clumpy and compact compared to N$_2$H$^+$ (1$-$0), HNC (1$-$0), and HCN (1$-$0), which have lower effective critical densities (1.0 $\times$ 10$^4$, 3.7 $\times$ 10$^3$, and 8.4 $\times$ 10$^3$ cm$^{-3}$, respectively).

\subsection{Fibers as Intermediate Structures Between Filaments and Cores}

The comparison between the observed line mass ($M_{\rm line}$) of fibers and the critical line mass ($M_{\rm crit} = 2\sigma^2_{\rm eff}/G$; \citealt{1964ApJ...140.1056O}) offers a key diagnostic for assessing their gravitational stability. Here, $\sigma_{\rm eff}$ represents the effective velocity dispersion that consists of thermal and nonthermal (turbulent) components, and $G$ is the gravitational constant. In environments where both thermal pressure and turbulent motions provide support against gravitational collapse, the effective dispersion is expressed as $\sigma_{\rm eff} = \sqrt{\sigma^2_{\rm nth} + c_s^2}$, where $\sigma_{\rm nth}$ is the nonthermal velocity dispersion and $c_s$ is the thermal sound speed. Accordingly, the critical line mass becomes a function of both temperature and turbulence \citep{2018A&A...610A..77H}:
\begin{equation}\label{fib_wid}
M_{\rm crit}(T, \sigma_{nth})=\frac{2c_s^2}{G}(1+(\frac{\sigma_{nth}}{c_s})^2) ,\\
\end{equation}
Considering the H$^{13}$CO$^+$ fibers as velocity-coherent substructures embedded within the filament, our analysis reveals that approximately 68\% exhibit line-mass ratios in the range $0.5 \leq M_{\rm line}/M_{\rm crit} \leq 1.5$, suggesting that the majority of fibers are gravitationally bounded or marginally stable. These fibers are likely in or near a state of quasi-equilibrium, which is a potential condition for further gravitational collapse \citep{2010ApJ...719L.185J}. In contrast, the remaining $\sim$32\% of fibers have $M_{\rm line}/M_{\rm crit} \leq 0.5$, indicating that they are subcritical and thus may represent transient features, potentially shaped by turbulence or external compression, but unlikely to form stars unless further mass accumulation occurs.

As discussed in Section~\ref{sec:res}, kinematic analysis of the DR21SF region reveals that more than 90\% of the area exhibits subsonic or transonic nonthermal velocity dispersions. This confirms that DR21SF is a dynamically quiescent filament on a spatial scale of $\sim$3.6 pc. In addition, based on the continuum emission and assuming a temperature of 15 K, we estimate the mass of DR21SF to be 110.2 $M_\odot$ using Equation~\ref{equ:gas_mass}, corresponding to a line mass of 30.6 $M_\odot$pc$^{-1}$. This value is approximately 64\% of the critical line mass, assuming a median Mach number of 0.97 at the same temperature. Although the total mass may be underestimated due to spatial filtering effects, this result may still suggest that the filament as a whole is gravitationally bound. At the smaller scale of dense cores (on the order of $\sim$0.01 pc), particularly the NH$2$D cores associated with dust continuum emission, six out of eight cores have virial parameters $\alpha_{\rm vir} < 2$, further confirming their gravitationally bound nature.

Taken together, these results reveal a complex internal substructure within the DR21SF filament, where multiple velocity-coherent fibers are embedded. Such a configuration is consistent with previous observational studies \citep{2013A&A...554A..55H,2015A&A...574A.104T}, suggesting that fibers represent the high-density, velocity-coherent components within filaments. Our observations indicate that quiescent filaments can contain gravitationally bound or quasi-stable fibers, which may play a key role in mediating mass accretion and setting the initial conditions for subsequent core formation.


\section{Summary} \label{Sum}

We conducted high-resolution NOEMA observations of DR21SF, a 3.6-pc-long, massive, and relatively quiescent filament in the Cygnus-X star-forming complex. Our main results are summarized as follows:

1. The 3~mm continuum and molecular lines emissions show good spatial agreement with the filament traced by the \textit{Herschel} column density map. While the H$^{13}$CO$^+$ emission appears more clumpy and confined to smaller regions, the emissions from other tracers, including continuum, CS and HCO$^+$, exhibit continuous and diffuse structures.
Mach number analysis indicates that the filament prominently exhibits subsonic or transonic nonthermal motions. The measured FWHMs of the filaments in continuum, HCO$^+$, and CS are 0.054, 0.029, and 0.030 pc, respectively.

2. A total of 32, 34, and 22 velocity-coherent fibers are identified in CS, HCO$^+$, and H$^{13}$CO$^+$, respectively, using a friend-of-friend algorithm. These fibers have lengths ranging from 0.07–0.39 pc (CS), 0.07–0.42 pc (HCO$^+$), and 0.06–0.29 pc (H$^{13}$CO$^+$). H$^{13}$CO$^+$ fibers also exhibit higher and more broadly distributed mass-per-unit-length values compared to CS and HCO$^+$ fibers.

3. Thirteen cold cores are identified from NH$_2$D emission. Most NH$_2$D cores are spatially and kinematically associated with H$^{13}$CO$^+$ fibers. Among the cores associated with the 3~mm continuum, those with reasonably estimated masses are gravitationally bound according to virial analysis, suggesting a prestellar nature.

4. The observed differences among CS, HCO$^+$, and H$^{13}$CO$^+$ line emissions are likely attributed to differences in their effective critical densities, with H$^{13}$CO$^+$ tracing denser and more quiescent gas.

5. Line mass analysis suggests that the majority of fibers are gravitationally bound and dynamically stable. Together with the subsonic or transonic turbulence in the filament and the presence of gravitational bound dense cores, this implies that velocity-coherent fibers within the filament create favorable conditions for core formation.


\begin{acknowledgments}

This work is supported by the National Natural Science Foundation of China (NSFC) grants Nos. 12425304 and U1731237, and National Key R\&D Program of China Nos. 2022YFA1603103 and 2023YFA1608204. K.Q. acknowledges the science research grant from the China Manned Space Project. K.Y. acknowledges supports from the National Natural Science Foundation of China under Grant Number 12503031, the Postdoctoral Fellowship Program of CPSF under Grant Number GZC20252099, the Shanghai Post-doctoral Excellence Program (No. 2024379), the Natural Science Foundation of Shanghai (No. 25ZR1402267) and the Yangyang Development Fund. This work is based on observations carried out under project number S18AS with NOEMA. NOEMA is supported by INSU/CNRS (France), MPG (Germany) and IGN (Spain).

\end{acknowledgments}


\clearpage

\bibliography{dr21sf}{}
\bibliographystyle{aasjournalv7}


\clearpage

\begin{appendix}

\section{Width estimation of the filament} \label{app1}

\begin{figure}[h!]
   \includegraphics[width=60mm]{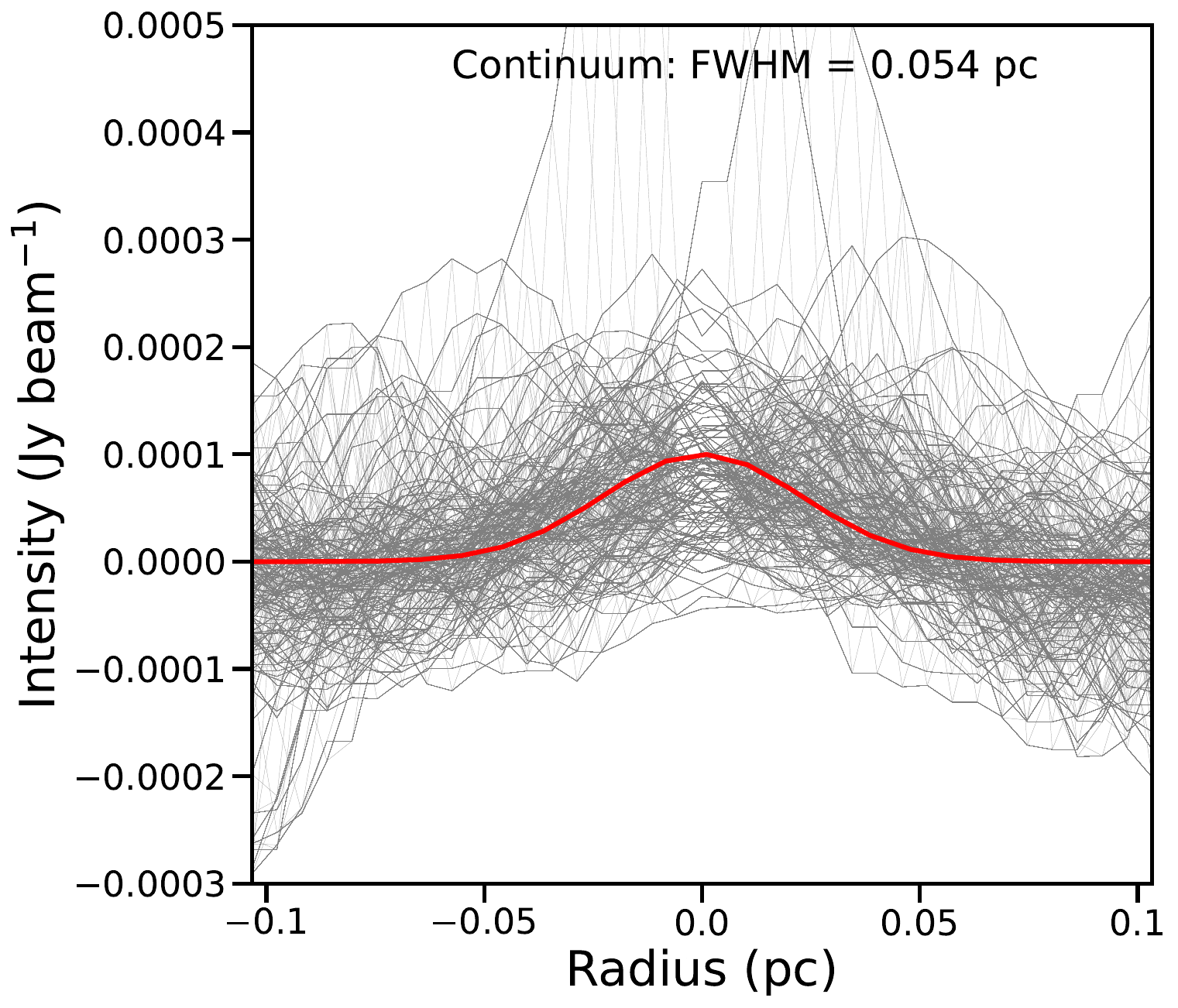}
   \includegraphics[width=57mm]{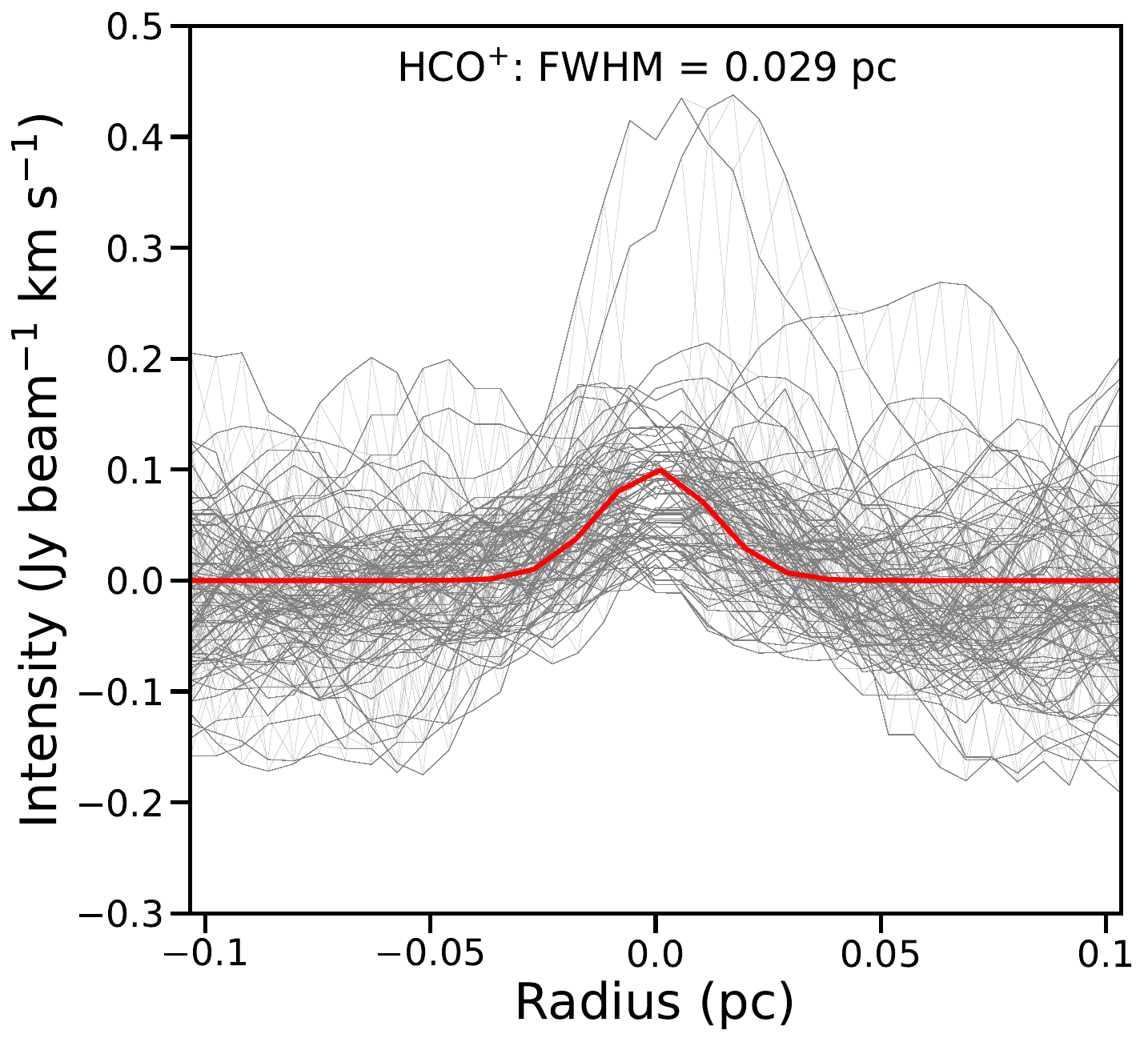}
   \includegraphics[width=57mm]{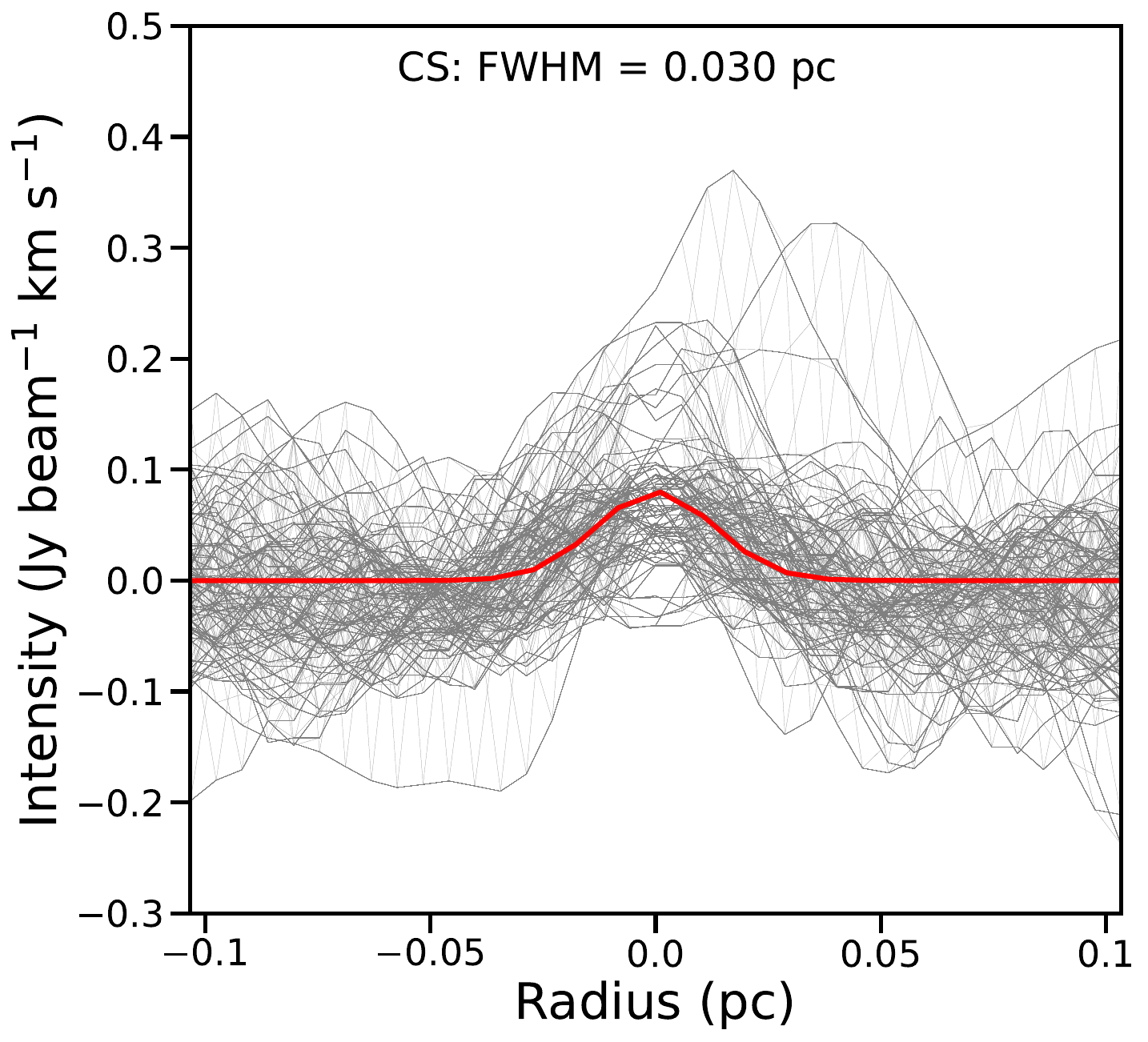}
     \caption{Radial intensity profile perpendicular to the continuum, HCO$^{+}$, and CS filaments. Individual integrated intensity profiles are shown in gray, and the red solid lines present the best-fit results of Gaussian fitting. The radius is the projected distance from the filament.}
  \label{fig:wid}
\end{figure}

To estimate the filament FWHM, we extracted perpendicular intensity profile at 6-pixel intervals (approximately one beam size linear scale; 1 pixel = 0.82$^{\prime\prime}$) along each filament and performed Gaussian fits to the median radial profiles. The Gaussian function used is:
\begin{equation}\label{fib_wid}
A(r) = A_{0}~{\rm exp}\left( \frac{-(r- \mu )^{2}}{2 \sigma^{2}_{\rm G}} \right) ,\\
\end{equation}
where $A(r)$ is the profile amplitude at the radial distance $r$, $A_{0}$ is the amplitude, $\mu$ is the mean, and $\sigma_{\rm G}$ is the standard deviation. The best-fit Gaussian profiles are shown as red solid lines in Figure \ref{fig:wid}. The derived FWHMs of the continuum, HCO$^{+}$, and CS filaments are 0.054, 0.029, and 0.030 pc, respectively.

\section{Properties of the identified fibers} \label{app2}

Table \ref{tab:prop_fibers_h13cop}-\ref{tab:prop_fibers_cs} present the properties of the identified fibers in different tracers. And Table \ref{tab:fiber_match} lists the matched fiber groups obtained by comparing the fibers identified in H$^{13}$CO$^{+}$, HCO$^{+}$, and CS. Fibers from different tracers are assigned to the same group when they exhibit spatial coincidence and similar velocity centroids. 

\begin{table*}[h!]
\centering
\movetableright=-0.5in
\caption{Properties of the H$^{13}$CO$^{+}$ fibers.}  \label{tab:prop_fibers_h13cop}
\begin{tabular}{c ccrc}
\hline\hline
 \multicolumn{5}{c}{Fibers in H$^{13}$CO$^{+}$} \\
\cline{1-5}
No. & $L$ & $W$ & $M$ & $M_{\rm line}$ \\
    & (pc) & (pc) & ($M_\odot$) & ($M_\odot$/pc) \\   
 (1) & (2) & (3) & (4) & (5) \\   
\hline
1  & 0.21 & 0.034 & 11  & 51 \\ 
2  & 0.26 & 0.034 & 11  & 42 \\ 
3  & 0.08 & 0.023 & 1.6 & 24 \\ 
4  & 0.16 & 0.031 & 5.7 & 37 \\ 
5  & 0.06 & 0.016 & 1.4 & 24 \\ 
6  & 0.10 & 0.026 & 1.6 & 16 \\ 
7  & 0.22 & 0.041 & 13  & 58 \\ 
8  & 0.13 & 0.035 & 3.3 & 26 \\ 
9  & 0.12 & 0.034 & 4.1 & 34 \\ 
10 & 0.08 & 0.023 & 5.0 & 66 \\ 
11 & 0.10 & 0.029 & 5.0 & 55 \\ 
12 & 0.29 & 0.046 & 25  & 85 \\ 
13 & 0.24 & 0.044 & 14  & 59 \\ 
14 & 0.06 & 0.018 & 1.5 & 24 \\ 
15 & 0.20 & 0.031 & 2.4 & 12 \\ 
16 & 0.27 & 0.039 & 13  & 47 \\ 
17 & 0.16 & 0.037 & 6.1 & 39 \\ 
18 & 0.12 & 0.017 & 2.7 & 23 \\ 
19 & 0.09 & 0.029 & 4.2 & 47 \\ 
20 & 0.11 & 0.024 & 6.6 & 62 \\
21 & 0.10 & 0.019 & 5.1 & 52 \\ 
22 & 0.11 & 0.022 & 4.4 & 39 \\ 
\hline
maximum & 0.29 & 0.046 & 25  & 85 \\
minimum & 0.06 & 0.016 & 1.5 & 12 \\
mean & 0.15 & 0.029 & 6.7 & 42 \\
median  & 0.12 & 0.030 & 5.0 & 41 \\
\hline\hline
\end{tabular}\\
  Note. (1): Numbers of the fibers. (2): Fiber length. (3): Fiber width. (4): Fiber mass. (5): Fiber mass-per-unit-length.
\end{table*}

\begin{table*}[h!]
\centering
\movetableright=-0.5in
\caption{Properties of the HCO$^{+}$ fibers.}  \label{tab:prop_fibers_hcop}
\begin{tabular}{c ccrc}
\hline\hline
 \multicolumn{5}{c}{Fibers in HCO$^{+}$} \\
\cline{1-5}
No. & $L$ & $W$ & $M$ & $M_{\rm line}$ \\
    & (pc) & (pc) & ($M_\odot$) & ($M_\odot$/pc) \\   
 (1) & (2) & (3) & (4) & (5) \\   
\hline
1  & 0.42 & 0.034 & 0.81  & 1.9  \\
2  & 0.14 & 0.029 & 0.16  & 1.2  \\
3  & 0.23 & 0.031 & 0.17  & 0.76 \\
4  & 0.09 & 0.021 & 0.03 & 0.29 \\
5  & 0.17 & 0.026 & 0.07 & 0.41 \\
6  & 0.31 & 0.029 & 0.40  & 1.0  \\
7  & 0.13 & 0.028 & 0.08 & 0.64 \\
8  & 0.17 & 0.041 & 0.14  & 0.82 \\
9  & 0.09 & 0.023 & 0.03 & 0.36 \\
10 & 0.10 & 0.028 & 0.10  & 1.1  \\
11 & 0.21 & 0.034 & 0.18  & 0.83 \\
12 & 0.14 & 0.023 & 0.03 & 0.19 \\
13 & 0.14 & 0.022 & 0.03 & 0.20 \\
14 & 0.15 & 0.032 & 0.21  & 1.4  \\
15 & 0.08 & 0.021 & 0.02 & 0.26 \\
16 & 0.20 & 0.041 & 0.16  & 0.80 \\
17 & 0.12 & 0.038 & 0.10  & 0.85 \\
18 & 0.14 & 0.026 & 0.03 & 0.19 \\
19 & 0.10 & 0.024 & 0.04 & 0.46 \\
20 & 0.12 & 0.019 & 0.12  & 0.98 \\
21 & 0.13 & 0.038 & 0.22  & 1.7  \\
22 & 0.10 & 0.021 & 0.03 & 0.34 \\
23 & 0.12 & 0.022 & 0.09 & 0.70 \\
24 & 0.11 & 0.028 & 0.06 & 0.56 \\
25 & 0.10 & 0.030 & 0.04 & 0.42 \\
26 & 0.09 & 0.023 & 0.02 & 0.27 \\
27 & 0.08 & 0.021 & 0.02 & 0.34 \\
28 & 0.08 & 0.025 & 0.06 & 0.78 \\
29 & 0.15 & 0.038 & 0.05 & 0.37 \\
30 & 0.15 & 0.026 & 0.10  & 0.69 \\
31 & 0.12 & 0.031 & 0.04 & 0.29 \\
32 & 0.08 & 0.020 & 0.04 & 0.51 \\
33 & 0.11 & 0.024 & 0.03 & 0.26 \\
34 & 0.11 & 0.032 & 0.11  & 1.0  \\
\hline
maximum & 0.42 & 0.041 & 0.81  & 1.9  \\
minimum & 0.08 & 0.019 & 0.02 & 0.19 \\
mean & 0.14 & 0.028 & 0.11  & 0.67 \\
median  & 0.12 & 0.027 & 0.07 & 0.60 \\
\hline\hline
\end{tabular}\\
\end{table*}

\begin{table*}[h!]
\centering
\movetableright=-0.5in
\caption{Properties of the CS fibers.}  \label{tab:prop_fibers_cs}
\begin{tabular}{c ccrc}
\hline\hline
 \multicolumn{5}{c}{Fibers in CS} \\
\cline{1-5}
No. & $L$ & $W$ & $M$ & $M_{\rm line}$ \\
    & (pc) & (pc) & ($M_\odot$) & ($M_\odot$/pc) \\   
 (1) & (2) & (3) & (4) & (5) \\   
\hline
1  & 0.39 & 0.048 & 11 & 30 \\ 
2  & 0.24 & 0.036 & 0.21 & 0.86 \\ 
3  & 0.11 & 0.037 & 1.2 & 11 \\ 
4  & 0.18 & 0.040 & 0.35 & 2.0 \\ 
5  & 0.36 & 0.032 & 4.8 & 13 \\ 
6  & 0.12 & 0.025 & 1.1 & 8.9 \\ 
7  & 0.32 & 0.021 & 1.8 & 5.6 \\ 
8  & 0.13 & 0.027 & 0.88 & 6.8 \\ 
9  & 0.11 & 0.031 & 1.1 & 9.6 \\ 
10 & 0.09 & 0.030 & 0.56 & 6.1 \\ 
11 & 0.15 & 0.032 & 1.7 & 11 \\ 
12 & 0.20 & 0.027 & 0.29 & 1.5 \\ 
13 & 0.12 & 0.029 & 0.50 & 4.1 \\ 
14 & 0.11 & 0.019 & 0.99 & 9.4 \\ 
15 & 0.10 & 0.028 & 0.30 & 3.0 \\ 
16 & 0.14 & 0.026 & 0.95 & 6.6 \\ 
17 & 0.12 & 0.024 & 0.73 & 6.1 \\ 
18 & 0.20 & 0.026 & 1.3 & 6.6 \\ 
19 & 0.08 & 0.025 & 0.33 & 4.0 \\ 
20 & 0.21 & 0.027 & 2.1 & 10 \\
21 & 0.21 & 0.024 & 1.9 & 9.0 \\ 
22 & 0.11 & 0.035 & 1.3 & 13 \\ 
23 & 0.13 & 0.030 & 1.3 & 10 \\ 
24 & 0.16 & 0.034 & 2.4 & 15 \\ 
25 & 0.29 & 0.030 & 0.65 & 2.2 \\ 
26 & 0.16 & 0.033 & 0.90 & 5.6 \\ 
27 & 0.17 & 0.032 & 1.1 & 6.4 \\ 
28 & 0.29 & 0.030 & 1.1 & 3.8 \\ 
29 & 0.19 & 0.049 & 2.2 & 12 \\ 
30 & 0.08 & 0.017 & 0.42 & 5.8 \\
31 & 0.09 & 0.022 & 0.36 & 4.4 \\ 
32 & 0.17 & 0.029 & 0.54 & 3.2 \\ 
\hline
maximum & 0.39 & 0.049 & 11 & 30 \\ 
minimum & 0.08 & 0.017 & 0.21 & 0.86 \\ 
mean & 0.17 & 0.030 & 1.5 & 7.7 \\ 
median  & 0.15 & 0.029 & 1.0 & 6.5 \\ 
\hline\hline
\end{tabular}\\
\end{table*}

\begin{table*}[h!]
\centering
\caption{Matched fiber groups across the three molecular tracers.}  \label{tab:fiber_match}
\begin{tabular}{c c c c}
\hline\hline
 Matched Set & H$^{13}$CO$^{+}$ & HCO$^{+}$ & CS \\
 & No. & No. & No. \\
\hline
1 & 1,2 & 1,4 & 1 \\
\hline
2 &  & 3 & 6 \\
\hline
3 & 3 & 6 & 7,9,11 \\
\hline
4 &  & 7 & 8 \\
\hline
5 &  & 8 & 5 \\
\hline
6 &  & 10 & 10 \\
\hline
7 &  & 11 & 13 \\
\hline
8 &  & 12 & 14 \\
\hline
9 &  & 14 & 16 \\
\hline
10 &  & 15,17 & 18 \\
\hline
11 &  & 16 & 19 \\
\hline
12 & 8 &  & 20 \\
\hline
13 &  & 20 & 21 \\
\hline
14 & 9 & 21 & 22 \\
\hline
15 &  & 23 & 24 \\
\hline
16 &  & 26 & 27 \\
\hline
17 & 12 & 27,28,29 &  \\
\hline
18 & 13 &  & 28,29 \\
\hline
19 & 15 & 30 &  \\
\hline\hline
\end{tabular}\\
\end{table*}

\end{appendix}


\end{document}